\documentclass[journal]{IEEEtran}
\usepackage[compatibility=false]{caption}
\IEEEoverridecommandlockouts
\usepackage{cite}
\usepackage{hyperref}
\usepackage{tikz}
\usepackage{amsmath,amssymb,amsfonts,mathtools,bm}
\usepackage{amsthm}
\usepackage{graphicx}
\usepackage{textcomp}
\usepackage{xcolor}
\usepackage{booktabs}
\usepackage{multirow}
\usepackage{subcaption}
\usepackage{algorithm}
\usepackage{algpseudocode}
\usepackage{url}
\usepackage{balance}
\usepackage{pifont}
\usepackage{array}
\usepackage{adjustbox}
\usepackage{makecell}
\usepackage{tabularx}

\definecolor{ourred}{RGB}{200,0,0}
\newcommand{\ourmodelval}[1]{\textcolor{ourred}{\textbf{#1}}}
\definecolor{ourblue}{RGB}{0,0,150}
\newcommand{\secondbestval}[1]{\textcolor{ourblue}{\underline{#1}}}

\newcommand{\cmark}{\textbf{\ding{51}}}
\newcommand{\xmark}{\textbf{\ding{55}}}

\DeclareRobustCommand{\orcidicon}{%
    \begin{tikzpicture}
    \draw[lime, fill=lime] (0,0) 
    circle [radius=0.16] 
    node[white] {{\fontfamily{qag}\selectfont \tiny ID}};     \draw[white, fill=white] (-0.0625,0.095) 
    circle [radius=0.007];     \end{tikzpicture}
    \hspace{-2mm}}
\foreach \x in {A, ..., Z}{%
    \expandafter\xdef\csname orcid\x\endcsname{\noexpand\href{https://orcid.org/\csname orcidauthor\x\endcsname}{\noexpand\orcidicon}}
    }

\allowdisplaybreaks
\renewcommand{\baselinestretch}{1}

\begin{document}

\title{RadioMapMotion: A Dataset and Baseline for Proactive Spatio-Temporal Radio Environment Prediction}

\author{
Honggang Jia\orcidA{},~\IEEEmembership{Student Member,~IEEE,}
Nan Cheng\orcidB{},~\IEEEmembership{Senior Member,~IEEE,}
Xiucheng Wang\orcidC{},~\IEEEmembership{Graduate Student Member,~IEEE,}

\thanks{ }
\thanks{
    }
}
\maketitle

\begin{abstract}
Radio maps (RMs), which provide location-based pathloss estimations, are fundamental to enabling proactive, environment-aware communication in 6G networks. However, existing deep learning-based methods for RM construction often model dynamic environments as a series of independent static snapshots, thereby omitting the temporal continuity inherent in signal propagation changes caused by the motion of dynamic entities. To address this limitation, we propose the task of spatio-temporal RM prediction, which involves forecasting a sequence of future maps from historical observations. A key barrier to this predictive approach has been the lack of datasets capturing continuous environmental evolution. To fill this gap, we introduce RadioMapMotion, the first large-scale public dataset of continuous RM sequences generated from physically consistent vehicle trajectories. As a baseline for this task, we propose RadioLSTM, a UNet architecture based on Convolutional Long Short-Term Memory (ConvLSTM) and designed for multi-step sequence forecasting. Experimental evaluations show that RadioLSTM achieves higher prediction accuracy and structural fidelity compared to representative baseline methods. Furthermore, the model exhibits a low inference latency, indicating its potential suitability for real-time network operations. Our project will be publicly released at: \url{https://github.com/UNIC-Lab/RadioMapMotion} upon paper acceptance.

\end{abstract}


\begin{figure*}[t]
    \centering
    \includegraphics[width=0.99\textwidth]{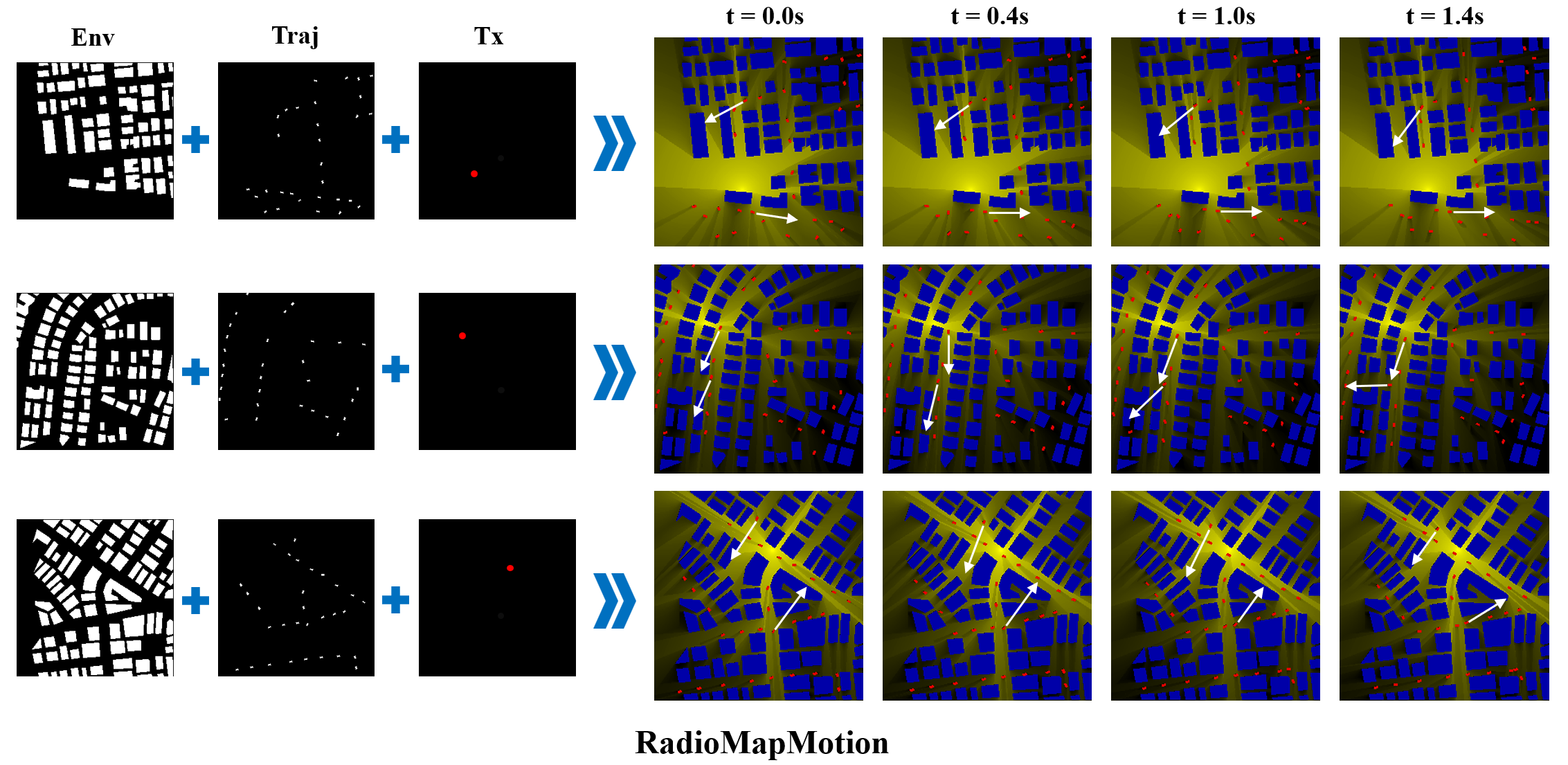}
    \captionsetup{font=small}
    \caption{Conceptual illustration of the structure of the RadioMapMotion dataset. Static environmental layouts (Env), dynamic vehicle trajectories (Traj), and transmitter locations (Tx) are combined to generate continuous radio map sequences, where signal propagation evolves causally over time in response to vehicle motion.}
    \label{fig:dataset_structure}
    \vspace{-10pt}
\end{figure*}

\section{Introduction}
\label{sec:introduction}
The sixth-generation (6G) wireless networks are expected to enable network management functions that operate with higher levels of automation and embedded intelligence~\cite{dang2020should, 9598915, sun2025comprehensive, sun2024knowledge}. Realizing this vision relies heavily on the digital twin (DT)~\cite{almasan2022network}, which involves creating a high-fidelity, virtual replica of the physical network and its surrounding environment, including dynamic vehicle-to-everything (V2X) communication scenarios~\cite{garcia2021tutorial, tao2022digital, lin20236g, 3}. To function effectively, the DT must maintain a precise and continuously updated representation of real-world conditions, particularly the complex and dynamic radio propagation environment~\cite{wu2021digital,ansari2025toward}. Accurate knowledge of pathloss is a fundamental component of this environmental awareness~\cite{letaief2019roadmap}, essential for optimizing resource allocation, interference management, and overall quality-of-service (QoS)~\cite{de2023qos}. Consequently, radio map (RM) technology, which provides a spatial representation of pathloss, has become a key component of DT~\cite{jiang2024physics, bi2019engineering,romero2022radio}. RMs offer the potential to acquire this critical information based on location alone, thereby reducing the reliance on continuous, overhead-intensive pilot transmissions, an issue especially challenging in scenarios involving devices like reconfigurable intelligent surfaces~\cite{liu2021reconfigurable}.

Traditionally, RMs have been constructed through two primary approaches: sampling-based methods that rely on spatial measurements within the target region, and sampling-free methods, such as those that utilize environmental modeling and electromagnetic ray tracing~\cite{mao2018constructing, cover1967nearest, breidt2000local}. While sampling-based methods face challenges related to measurement density and coverage limitations, ray-tracing approaches suffer from high computational complexity that impedes real-time implementation\cite{fuschini2015ray}. To overcome these limitations, neural network-based approaches have emerged as promising alternatives~\cite{wang2025radiodiff_3d, wang2025radiodiff_inverse, zhang2023rme}, with architectures such as RadioUNet demonstrating the feasibility of efficiently constructing high-fidelity RMs from environmental inputs~\cite{levie2021radiounet}.
 
Despite these advances, current radio map construction methodologies still struggle to support real-time operation in highly dynamic scenarios. Although recent efforts have extended RM modeling to account for dynamic elements by incorporating vehicle positions as conditional inputs~\cite{wang2025radiodiff}, these approaches typically treat each configuration as an independent static snapshot and often incur significant inference latency, making them ill-suited for latency-critical applications such as autonomous driving or V2X services. More fundamentally, even the most recent advances, including diffusion-based generative models~\cite{wang2025radiodiff, wang2025radiodiff_k2}, remain focused on reconstructing the current electromagnetic state rather than forecasting its future evolution. As illustrated by widely used datasets such as RadioMapSeer~\cite{levie2021radiounet} and some 3D extensions~\cite{yapar2022dataset,wang2025radiodiff_3d}, RMs are generally generated under the assumption that vehicles remain stationary during measurement or simulation. While this simplification enables tractable modeling, it inherently fails to capture the continuous, causal relationship between mobile obstacles and signal variation over time. Consequently, current methods, regardless of their reconstruction fidelity or latency, are reactive by design and lack the capability for proactive prediction. This limitation can lead to communication disruptions that compromise both safety and performance in high-mobility vehicular networks. Introducing a predictive time margin would enable anticipatory scheduling or connection reassignment, thereby mitigating such disruptions and enhancing communication reliability.

This paper formulates the task of spatio-temporal RM prediction: forecasting a time-ordered sequence of future RMs from historical observations. Unlike methods that generate a map from a static snapshot, this formulation explicitly accounts for the continuous evolution of signal propagation in response to moving vehicles. It allows networks to base decisions on anticipated future states, rather than current or past measurements alone. To support progress in this emerging field, we present RadioMapMotion, the first large-scale dataset explicitly designed to capture the temporally continuous evolution of radio environments. Existing datasets consist of isolated snapshots, RadioMapMotion provides continuous sequences where radio variations are driven by vehicle motion. For each scenario, we simulate realistic vehicular trajectories at fine-grained time intervals and generate corresponding RMs that reflect the time-varying impact of mobility on wireless propagation.

Building on this foundation, we propose RadioLSTM, a ConvLSTM-based UNet architecture~\cite{shi2015convolutional,zheng2024precipitation} specifically suited for spatio-temporal RM prediction. In contrast to frame-by-frame reconstruction methods, RadioLSTM explicitly learns the underlying spatio-temporal dynamics from historical RM sequences to forecast future propagation states. The model integrates the spatio-temporal modeling capacity of ConvLSTM within a UNet framework, enabling it to preserve spatial structures while effectively capturing temporal dependencies.

The primary contributions of this work are as follows:
\begin{enumerate}
    \item We introduce RadioMapMotion, the first large-scale dataset designed to capture the continuous evolution of radio environments over time. Unlike prior datasets composed of isolated snapshots, RadioMapMotion preserves the physically grounded causal relationship between vehicle motion and signal variation, establishing a solid foundation for learning-based predictive modeling.
    
    \item Building upon this dataset, we formally define the task of spatio-temporal RM prediction. This new formulation shifts the research objective from merely reconstructing the present state—a key limitation of current ``dynamic" approaches that treat mobility as static snapshots—to forecasting its future evolution, enabling proactive network operations.
    
    \item We propose RadioLSTM, a ConvLSTM-UNet hybrid architecture, as a strong and efficient baseline for this new task. By leveraging historical measurements, our model captures both spatial fidelity and temporal progression without requiring real-time environmental sensing.
    
    \item Through comprehensive experiments, we demonstrate that RadioLSTM delivers high prediction accuracy and structural fidelity. Its low inference latency further validates its suitability for real-time, proactive network management.
\end{enumerate}

The remainder of this paper is organized as follows. Section \ref{sec:related_work} reviews related work. Section \ref{sec:problem_formulation} formally defines the spatio-temporal RM prediction problem and justifies the availability of historical measurements. Section \ref{sec:dataset} introduces the RadioMapMotion dataset. Section \ref{sec:methodology} details the RadioLSTM architecture. Section \ref{sec:experiments} presents experimental results, and Section \ref{sec:conclusion} concludes the paper.

\begin{table*}[ht]
    \centering
    \captionsetup{font=small}
    \caption{Comparison of RadioMapMotion with Other Public RM Datasets.}
    \label{tab:dataset_comparison}
    \renewcommand{\arraystretch}{1.1}
    \begin{tabular}{l|c|cccc}
      \toprule
      \textbf{Properties} & \textbf{RadioMapMotion (Ours)} & \textbf{RadioMapSeer~\cite{levie2021radiounet}} & \textbf{UrbanRadio3D~\cite{wang2025radiodiff_3d}} & \textbf{SpectrumNet~\cite{SpectrumNet}} & \textbf{CKMImageNet~\cite{ckmimage}} \\
      \midrule
      \textbf{Total Frames} & \textbf{450K} & 56K & 11.2M & 300K & 72K \\
      \textbf{Number of Maps} & \textbf{300} & 701 & 701 & 764 & 42 \\
      \textbf{Number of BS Locations} & \textbf{20} & 80 & 200 & 4 & 1$\sim$42 \\
      \textbf{Map Resolution} & \textbf{256$\times$256} & 256$\times$256 & 256$\times$256 & 128$\times$128 & 128$\times$128 \\
      \textbf{Horizontal Resolution} & \textbf{1 m} & 1 m & 1 m & 10 m & 2 m \\
      \textbf{Raw Pathloss Data} & \cmark & \xmark & \xmark & \xmark & \xmark \\
      \textbf{Vehicle Trajectories} & \cmark & \xmark & \xmark & \xmark & \xmark \\
      \bottomrule
    \end{tabular}
\end{table*}

\section{Related Work}
\label{sec:related_work}
This section discusses related work in deep learning-based radio map construction, spatio-temporal prediction methods adapted from video forecasting, and AI-driven approaches to proactive network management, highlighting how our formulation builds upon and differs from these lines of research.

\begin{figure}[ht]
    \centering
    \captionsetup{font=small}
    \begin{subfigure}[b]{0.4\columnwidth}
      \centering
      \includegraphics[width=\textwidth]{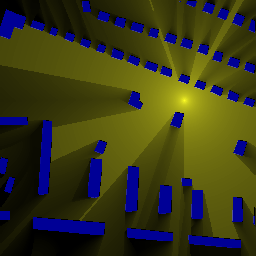} 
      \caption{Static Radio Map.}
      \label{fig:srm}
    \end{subfigure}
    \begin{subfigure}[b]{0.4\columnwidth}
      \centering
      \includegraphics[width=\textwidth]{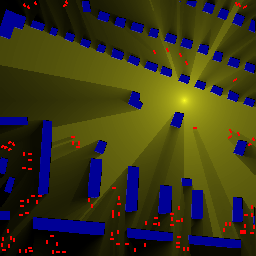} 
      \caption{Dynamic Radio Map.}
      \label{fig:drm}
    \end{subfigure}
    \caption{
      Illustration of the SRM versus the DRM. The yellow heatmap represents pathloss intensity. Blue elements are static buildings. The DRM in (b) additionally includes vehicles (represented as red elements).
    }
    \label{fig:srm_drm_radiodiff_comparison}
\end{figure}

\subsection{Deep Learning for Static RM Construction}
Deep learning has notably advanced RM construction, offering a cost-effective and computationally efficient alternative to sampling-based methods and ray-tracing. Early efforts adapted established computer vision architectures for this image-to-image translation task. The pioneering RadioUNet~\cite{levie2021radiounet} demonstrated the effectiveness of UNet in learning the mapping from environmental layouts to pathloss maps. Subsequent works explored generative adversarial networks, such as RME-GAN~\cite{zhang2023rme},recent state-of-the-art performance has been achieved by diffusion models, especially RadioDiff~\cite{wang2025radiodiff}. RadioDiff formulates RM construction as a conditional generative problem and leverages diffusion modeling to produce high-fidelity outputs, arguing that this better matches the generative nature of the task compared to earlier discriminative approaches.

These works commonly distinguish between static radio maps (SRM), which model only fixed obstacles such as buildings, and dynamic radio maps (DRM), which additionally account for mobile obstacles like vehicles. As illustrated in Fig.~\ref{fig:srm_drm_radiodiff_comparison}, which follows the visualization style of~\cite{wang2025radiodiff}, DRMs contain vehicle-induced pathloss patterns that do not appear in SRMs. However, a common characteristic of these approaches is that they treat each environmental state as an independent, static snapshot. We term this approach pseudo-dynamic because it considers the presence of vehicles but does not explicitly model their temporal motion or the causal evolution of the radio environment over time. This pseudo-dynamic approach cannot capture the continuous, causally linked relationship between vehicle movement and signal variation. Each DRM is generated independently, with no modeling of past or future states. As a result, these models are primarily suited for reconstruction rather than prediction, they can reconstruct the RM for a given static scene but cannot predict how it will change as vehicles move. This limitation impedes proactive network optimization in high-mobility scenarios, which motivates our proposed shift from static generation to spatio-temporal prediction.

\subsection{Spatio-Temporal Prediction and Video Forecasting}
Spatio-temporal prediction is a well-established challenge in AI, with applications spanning weather forecasting, traffic modeling, and beyond~\cite{amato2020novel}. Video prediction, a task of forecasting future frames from historical sequences, serves as a direct conceptual analog to our work~\cite{oprea2020review}. Early methods typically combined CNNs for spatial feature extraction with recurrent neural networks (RNNs) to model temporal dynamics~\cite{mathieu2015deep}. A foundational architecture in this domain is the Convolutional LSTM (ConvLSTM)~\cite{shi2015convolutional}, which replaces fully connected operations in standard LSTMs with convolutions, enabling spatially aware temporal modeling. ConvLSTM has since become widely adopted in video prediction tasks, notably in precipitation nowcasting. Our proposed RadioLSTM builds upon this framework, integrating it into a UNet structure~\cite{zheng2024precipitation} tailored for the spatial structure and resolution requirements of RM prediction.

More recent advances have introduced Transformer-based models and diffusion approaches for video forecasting~\cite{villegas2022phenaki, ho2022video}. Concurrently, State space models (SSMs), particularly the Mamba architecture~\cite{gu2023mamba}, have emerged as efficient alternatives for long-sequence modeling. Mamba combines the linear-time inference of RNNs with the parallel training capability of Transformers, showing promise on extended temporal sequences. Given its efficiency and scalability, Mamba represents a compelling modern baseline for comparison with recurrent architectures in spatio-temporal RM prediction. In our work, we design a hybrid ConvMambaUNet that integrates spatial convolutions with Mamba’s sequence modeling to preserve fine-grained radio map structures while capturing temporal dynamics.

\subsection{AI-Enabled Proactive Network Management}
The concept of using prediction to support anticipatory control has been explored in the wireless community, with the goal of moving network management beyond purely reactive strategies. This paradigm has been investigated across various critical network functions, yet existing works differ fundamentally from our approach in terms of their prediction target's scope and dimensionality. For instance, in mobility management, researchers have used deep learning to predict user trajectories, which enables proactive handover decisions that reduce service interruptions~\cite{lima2023deep}. This approach is particularly vital in millimeter-wave (mmWave) and terahertz (THz) systems, where predictive beam management, which forecasts the optimal beam alignment based on user movement or channel evolution, is essential for circumventing link blockages and avoiding the high overhead of reactive beam searching~\cite{xue2024survey}. Other works focus on forecasting future channel state information (CSI) or QoS requirements to facilitate proactive resource allocation. For example,~\cite{zhou2024transformer} successfully applied a Transformer network to predict a future sequence of CSI values based on historical data in high-speed railway scenarios, demonstrating a powerful method to counteract the effects of channel aging. While such CSI prediction methods may also treat channel data as a sequence of two-dimensional inputs, the ``image" in that context represents the channel's response in the time-frequency domain for a single communication link, rather than the pathloss distribution across a geographical area~\cite{evoCSINet}. Similarly, some research has addressed the spatio-temporal prediction of spectrum maps to aid dynamic spectrum access~\cite{10906611}. These efforts, however, typically rely on interpolating from sparse sensor measurements and predict spectrum occupancy, which is a stochastic process driven by user activity, not the physically determined pathloss governed by environmental geometry. A closer conceptual parallel involves modeling the dynamic RM evolution by interpreting it as a spatio-temporal video stream~\cite{DMD}. Although this work shares a similar problem conceptualization, its proposed solution is based on Dynamic Mode Decomposition, a classical method that is constrained by its underlying assumption of linearity and may not fully capture the complex, non-linear dynamics of radio propagation in cluttered environments. In contrast, our work is the first to systematically formulate and solve this problem using an advanced, end-to-end deep learning paradigm adapted from video forecasting. Our work differs in its objective to predict a high-dimensional, spatially-extended representation, the entire RM. This approach offers a more complete characterization of the future radio environment, potentially enabling network-wide optimizations that are more challenging to achieve with single-point or single-link predictions.

\section{Problem Formulation}
\label{sec:problem_formulation}
In this section, we formalize the task of spatio-temporal RM prediction. The goal is to forecast a sequence of future RMs based on historical observations, rather than reconstructing a single static map.

\subsection{Spatio-Temporal Radio Environment}
We consider a dynamic environment where the radio propagation characteristics evolve over time due to the movement of objects such as vehicles. The state of the radio environment at any discrete time step $t$ is represented by a RM $\mathbf{P}_t \in \mathbb{R}^{N \times N}$, where each element represents the pathloss at a specific grid location. A sequence of these maps over a time interval constitutes a radio environment ``video''.
The input to our prediction model is a sequence of $T_c$ historically observed RMs, which we term the context sequence as follows.
\begin{equation}
    \mathcal{X} = (\mathbf{P}_{t-T_c+1}, \mathbf{P}_{t-T_c+2}, \dots, \mathbf{P}_t).
\end{equation}
The model's objective is to predict the sequence of the next $T_p$ RMs, termed the target sequence as follows.
\begin{equation}
    \mathcal{Y} = (\mathbf{P}_{t+1}, \mathbf{P}_{t+2}, \dots, \mathbf{P}_{t+T_p}).
\end{equation}

\subsection{Justification of the Observational Premise}
A core premise of our formulation is the availability of the historical context sequence $\mathcal{X}$ at the network side. This assumption is grounded in the operational reality of modern wireless systems, where time-series channel measurements (e.g., CSI, RSRP) are routinely collected from user equipment or infrastructure sensors. In principle, these raw measurements can be converted into dense radio maps using established deep learning-based static RM reconstruction models. However, it is important to note that such reconstructed maps may contain noise or bias due to imperfect channel estimation or model generalization errors. In this work, our experiments assume access to ground-truth historical RMs, as provided by the simulation-based RadioMapMotion dataset, to isolate the performance of the predictive model itself. The practical deployment of such a system would require either highly accurate RM reconstruction or an end-to-end framework that jointly optimizes reconstruction and prediction, topics we identify as important directions for future work.

This approach offers two key advantages. First, it aligns with the causal nature of prediction: future states are inferred directly from past observations. Second, it reduces reliance on external ``God's-eye view" information, such as the precise real-time location of every vehicle. Instead, the model implicitly learns the effects of environmental dynamics through their observable impact on RMs, which may reduce the dependency on real-time sensing for deployment.

\subsection{Formal Problem Statement}
The task of spatio-temporal RM prediction is to learn a function $\mathcal{F}_{\theta}$, parameterized by weights $\theta$, that maps a context sequence $\mathcal{X}$ to a predicted future sequence $\hat{\mathcal{Y}}$.

\noindent\textbf{Problem 1. Spatio-Temporal Radio Map Prediction}
Given a dataset $\mathcal{D} = \{(\mathcal{X}^{(i)}, \mathcal{Y}^{(i)})\}_{i=1}^{M}$ consisting of $M$ context-target sequence pairs, the goal is to find the optimal parameters $\theta^*$ for a predictive model $\mathcal{F}_{\theta}$ that minimizes a loss function $\mathcal{L}(\cdot, \cdot)$ over the entire sequence as follows.
\begin{equation}
    \theta^* = \arg\min_{\theta} \mathbb{E}_{(\mathcal{X}, \mathcal{Y}) \sim \mathcal{D}} \left[ \mathcal{L}(\mathcal{F}_{\theta}(\mathcal{X}), \mathcal{Y}) \right].
\end{equation}
Here, $\hat{\mathcal{Y}} = (\hat{\mathbf{P}}_{t+1}, \dots, \hat{\mathbf{P}}_{t+T_p})$ denotes the predicted sequence of future RMs. This formulation aligns the problem with the established framework of sequence-to-sequence video forecasting.
To train and evaluate models for the task defined in Problem 1, a dataset must provide continuous, temporally coherent sequences of RMs that preserve physical causality. However, existing public datasets for RM research, such as the widely used RadioMapSeer benchmark, are designed primarily for static generation. They comprise independent environmental snapshots, lacking temporal continuity and causal evolution between frames. This gap limits their applicability for predictive modeling. Consequently, advancing research in predictive radio environment modeling requires a purpose-built dataset. This necessity motivates the development of the dataset introduced in the next section.

\section{The RadioMapMotion Dataset}
\label{sec:dataset}
To bridge the research gap in spatio-temporal radio environment prediction, we introduce \textit{RadioMapMotion}, a large-scale, high-fidelity dataset specifically designed to capture the continuous evolution of RMs in dynamic urban environments. This section outlines its key features, construction pipeline, and organizational structure.

\subsection{Key Features and Comparisons}
RadioMapMotion differs from existing datasets primarily in its underlying design philosophy. As summarized in Table~\ref{tab:dataset_comparison}, while prominent benchmarks have advanced the field, such as RadioMapSeer\cite{levie2021radiounet} and UrbanRadio3D\cite{wang2025radiodiff_3d} offering numerous static scenarios, SpectrumNet\cite{SpectrumNet} providing multi-band diversity, and CKMImageNet\cite{ckmimage} incorporating rich experimental details, they fundamentally lack the temporal continuity required for predictive modeling.

In contrast, the defining feature of RadioMapMotion is its inclusion of physically consistent vehicle trajectories. Each sample is a 15-frame video sequence in which the radio environment evolves causally in response to vehicle motion, precisely the structure needed to train and evaluate models defined in Section~\ref{sec:problem_formulation}. Moreover, RadioMapMotion is the only dataset in this comparison that provides raw, grid-based pathloss values in \texttt{.npz} format, enabling researchers to implement custom preprocessing and normalization beyond the provided grayscale images.

\subsection{Dataset Construction Pipeline}
RadioMapMotion was generated through a three-stage pipeline: dynamic environment preparation, ray-tracing simulation, and data post-processing.

\subsubsection{Dynamic Environment Preparation}
We augment static urban layouts from the RadioMapSeer dataset using a custom trajectory engine. Initially, vehicles are randomly seeded onto programmatically identified drivable areas, maintaining a minimum spacing of four vehicle lengths and adhering to a unified clockwise traffic flow to ensure organized, collision-free movement from the start. Vehicle motion at each time step is governed by a two-stage process.

First, for navigation, each vehicle evaluates potential forward paths by probing three primary directions: straight ahead and approximately $\pm45^{\circ}$. It identifies the nearest road pixel at a look-ahead distance of 1.5 vehicle lengths for each direction and selects the path most aligned with its current heading. This ensures logical forward motion. If a vehicle remains stationary for more than three frames, this mechanism adaptively expands its search to include wider $\pm90^{\circ}$ probes to find an alternative route. Second, directional smoothing ensures fluid turns. Instead of instantly changing orientation, a vehicle's new heading is calculated by interpolating between its current and target directions, using a smoothing factor of 0.4. This creates natural turning dynamics. A concurrent collision avoidance system prevents vehicles from intersecting with buildings or each other. Each simulated frame represents a 0.1-second interval, aligning the vehicle movements with realistic urban speeds and the 10-Hz reporting rate of V2X systems, such that a speed of 1 meter per frame corresponds to 36 km/h. We note that this simulation uses rule-based movements and has not been validated against real-world traffic data. Nevertheless, RadioMapMotion provides a crucial resource of temporally coherent data, establishing a foundation for the development and benchmarking of predictive radio environment models.

\subsubsection{Ray-Tracing Simulation}

\begin{table}[ht]
    \centering
    \captionsetup{font=small}
    \caption{Key Parameters for Ray-Tracing Simulation.}
    \label{tab:simulation_params}
    \renewcommand{\arraystretch}{1.1}
    \begin{tabularx}{0.75\linewidth}{lc}
      \toprule
      \textbf{Parameter} & \textbf{Value} \\
      \midrule
      Propagation model & Dominant Path Model \\
      Carrier frequency & 3.5 GHz \\
      Transmission power & 23 dBm \\
      Transmitter (TX) height & 1.5 m \\
      Receiver (RX) height & 1.5 m \\
      Simulation area & 256 m $\times$ 256 m \\
      Spatial resolution & 1 m \\
      Antenna type & Isotropic \\
      \bottomrule
    \end{tabularx}
\end{table}

With the dynamic 3D scenes prepared, we generated the corresponding pathloss data using Altair WinProp's high-fidelity ray-tracing engine. We employed the Dominant Path Model (DPM) for this task. The specific simulation parameters, chosen to reflect a realistic 6G V2X scenario, are detailed in Table~\ref{tab:simulation_params}. The DPM provides a suitable trade-off between computational efficiency and accuracy for generating a large-scale dataset. More exhaustive methods like Intelligent Ray Tracing (IRT) can capture a richer set of multipath components, their computational cost is prohibitive for simulating hundreds of thousands of unique, dynamic scenes. The DPM focuses on the most significant propagation paths, such as line-of-sight, single reflections, and diffractions, which are the primary contributors to the overall path loss in urban environments, thereby enabling the fast generation of accurate RMs.

\subsubsection{Data Post-processing}
The raw output from the WinProp simulator is processed through a standardized pipeline to generate two data formats: dense numerical arrays and 8-bit grayscale images. This ensures consistent, ready-to-use data for deep learning applications.

First, the sparse simulation output, which consists of receiver coordinates and corresponding pathloss values, is aggregated onto the $256 \times 256$ grid. For each grid location, the pathloss value $P_{\text{dB}}$ (in dB) is extracted and stored as a floating-point matrix. This full-resolution grid is saved as a compressed NumPy array \texttt{.npz}, forming the high-fidelity numerical format of the dataset. Second, for compatibility with vision-based models, we convert the numerical data into 8-bit grayscale \texttt{.png} images. The conversion for each value $P_{\text{dB}}$ at coordinate $(x, y)$ follows three steps.

\begin{itemize}
\item \textbf{Rasterization}: A $256 \times 256$ matrix $\mathbf{M}$ (initialized to zero) is populated by mapping continuous coordinates $(x, y)$ to discrete indices $(i, j)$ via ceiling as follows.
    \begin{equation}
      i = \lceil x \rceil - 1, \quad j = \lceil y \rceil - 1,
    \end{equation}
    where $i, j \in \{0, 1, \dots, 255\}$.

    \item \textbf{Clipping and Normalization}: $P_{\text{dB}}$ is clipped to $[-135, -39.5]$ dB, which are determined by applying the same procedure used in prior work to our dataset, then linearly scaled to $P_{\text{norm}} \in [0, 1]$ as follows.
    \begin{equation}
      P_{\text{norm}} = \frac{\max(P_{\min}, \min(P_{\max}, P_{\text{dB}})) - P_{\min}}{P_{\max} - P_{\min}}.
    \end{equation}
    Then assigned to $\mathbf{M}_{i,j}$.

    \item \textbf{Quantization to 8-bit}: The normalized matrix is scaled to 8-bit integer range as follows.
    \begin{equation}
      I_{i,j} = \lfloor \mathbf{M}_{i,j} \times 255 \rfloor.
    \end{equation}
    The resulting matrix $\mathbf{I}$ is saved as a grayscale \texttt{.png} image.
\end{itemize}

This pipeline produces spatially accurate and visually consistent representations of pathloss for training and evaluation.

\subsection{Dataset Structure}
\label{subsec:dataset_structure}
The RadioMapMotion dataset is organized hierarchically, as shown in Fig.~\ref{fig:dataset_structure}, to capture spatio-temporal dynamics in urban radio environments. It contains two parallel directories: one with raw pathloss data in \texttt{.npz} format, and the other with corresponding grayscale images in \texttt{.png} format. Both follow identical hierarchies to ensure pixel-perfect alignment between numerical and visual data. At the top level, the dataset includes 300 distinct urban environments. For each, we simulate 5 diverse vehicle trajectories, a number chosen to balance diversity and diminishing returns. For every trajectory, 20 fixed transmitter locations are evaluated, enabling models to learn location-dependent propagation patterns. Each environment-trajectory-transmitter combination yields a 15-frame sequence, capturing continuous channel evolution due to vehicle motion\footnote{
Naming convention: Directories are named \texttt{env\_XXX} (XXX: 000–299), \texttt{traj\_XX} (XX: 00–04), and \texttt{tx\_XX} (XX: 00–19). Each sequence contains 15 frames, indexed 00–14 in \texttt{.png}/\texttt{.npz} files.}. This structure results in $300 \times 5 \times 20 = 30,000$ sequences and $450,000$ individual maps. The organization supports efficient data loading and enables rigorous, multi-dimensional evaluation of model generalization, as described in Section~\ref{sec:experiments}.

\section{The Proposed RadioLSTM}
\label{sec:methodology}
To establish a strong baseline for the spatio-temporal RM prediction task, we propose RadioLSTM, an architecture specifically designed to model and forecast the evolution of radio environments. The core design principle of RadioLSTM is to integrate the proven spatial feature extraction capabilities of a UNet architecture with the temporal sequence modeling power of Convolutional LSTMs (ConvLSTMs). This hybrid approach addresses our problem by learning spatial patterns within each RM and capturing dynamics across the sequence. RadioLSTM learns the temporal evolution of the radio environment from historical observations, supporting a move from map reconstruction to prediction.

\subsection{Overall Architecture}
\begin{figure*}[t]
    \centering
    \captionsetup{font=small}
    \includegraphics[width=0.95\textwidth]{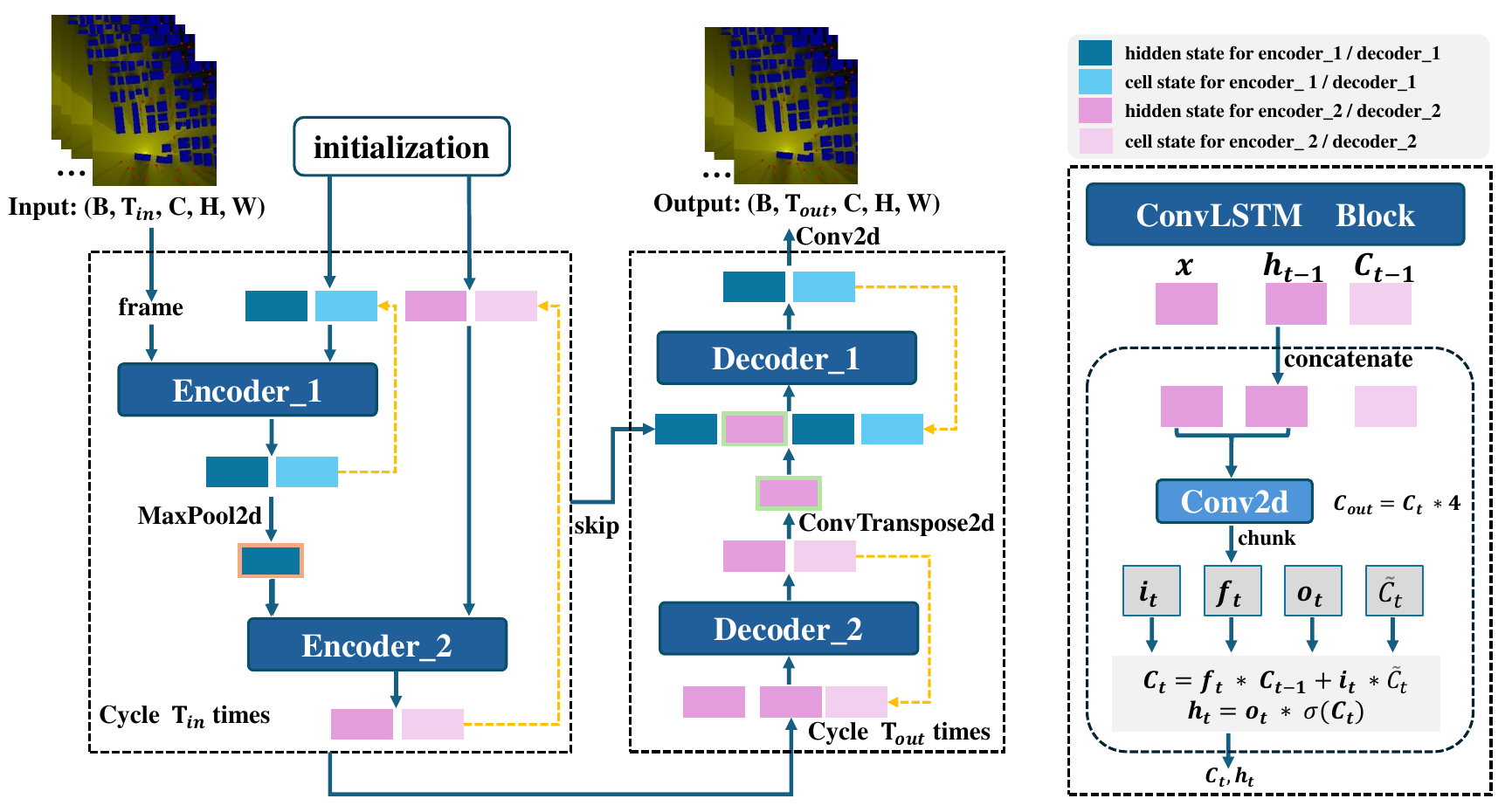} 
    \caption{The overall architecture of RadioLSTM. The model adopts a UNet-like encoder-decoder structure, where the core blocks in both the encoder and decoder are replaced by ConvLSTM cells to capture spatio-temporal dynamics. The right panel provides a detailed view of the internal structure of a single ConvLSTM block.}
    \label{fig:architecture}
    \vspace{-12pt}
\end{figure*}

RadioLSTM is implemented as a sequence-to-sequence model with a recurrent encoder-decoder structure, as depicted in Fig. \ref{fig:architecture}. The model takes a historical context sequence of $T_c$ RMs, $\mathcal{X} = (\mathbf{P}_{t-T_c+1}, \dots, \mathbf{P}_t)$, as input and is trained to generate a future target sequence of $T_p$ RMs, $\hat{\mathcal{Y}} = (\hat{\mathbf{P}}_{t+1}, \dots, \hat{\mathbf{P}}_{t+T_p})$.

The architecture is built upon a UNet-like skeleton, which is highly effective for image-to-image translation tasks due to its ability to preserve high-resolution spatial details through skip connections. However, unlike a standard UNet that uses simple convolutional layers, every feature extraction block in RadioLSTM is replaced with a ConvLSTM cell. This fundamental change transforms the network from a static image processor into a dynamic sequence processor, enabling it to learn the causal relationships between consecutive RMs. The model operates in two distinct phases: an \textit{encoding phase} that processes the entire context sequence to build a spatio-temporal memory, and a \textit{forecasting phase} that autoregressively generates the future sequence.

\subsection{ConvLSTM Cell}
The fundamental building block of RadioLSTM is the ConvLSTM cell ~\cite{shi2015convolutional}. A standard LSTM uses fully-connected operations for state transitions, which are effective for 1D sequences but do not preserve spatial relationships. In contrast, the ConvLSTM cell replaces these operations with convolutions, allowing it to handle 2D or 3D spatial data like our RMs.

A ConvLSTM cell maintains two internal states: a hidden state $\mathcal{H}_t$ and a cell state $\mathcal{C}_t$, both of which are 3D tensors preserving the spatial dimensions of the input. The flow of information is controlled by three gates: an input gate $i_t$, a forget gate $f_t$, and an output gate $o_t$, which are computed at each time step $t$ as follows.
\begin{align}
    i_t &= \sigma(W_{xi} * \mathbf{P}_t + W_{hi} * \mathcal{H}_{t-1} + b_i), \\
    f_t &= \sigma(W_{xf} * \mathbf{P}_t + W_{hf} * \mathcal{H}_{t-1} + b_f), \\
    o_t &= \sigma(W_{xo} * \mathbf{P}_t + W_{ho} * \mathcal{H}_{t-1} + b_o), \\
    g_t &= \tanh(W_{xg} * \mathbf{P}_t + W_{hg} * \mathcal{H}_{t-1} + b_g) ,\\
    \mathcal{C}_t &= f_t \odot \mathcal{C}_{t-1} + i_t \odot g_t ,\label{eq:cell_update}\\
    \mathcal{H}_t &= o_t \odot \tanh(\mathcal{C}_t), \label{eq:hidden_update}
\end{align}
where $*$ denotes the convolution operator and $\odot$ denotes the Hadamard product. The gates selectively determine what information to discard from the old cell state through the forget gate $f_t$ and what new information to store via the input gate $i_t$ and a candidate state $g_t$, as shown in Eq.~\ref{eq:cell_update}. The new hidden state $\mathcal{H}_t$ is then computed from the updated cell state, modulated by the output gate $o_t$, as given in Eq.~\ref{eq:hidden_update}. This mechanism allows the network to learn complex temporal dynamics while maintaining spatial awareness.

\subsection{Detailed Network Structure}
The RadioLSTM architecture consists of a two-level encoder and a corresponding two-level decoder, forming a recurrent UNet.

\subsubsection{Spatio-Temporal Encoder}
The encoder processes the input context sequence $\mathcal{X}$ frame by frame for $T_c$ time steps. It consists of two hierarchical levels. The first level takes an input frame $\mathbf{P}_t$ and passes it through a ConvLSTM cell, producing a hidden state $\mathcal{H}_t^1$. This hidden state is then spatially downsampled by a factor of two using a max-pooling layer. The result is fed into the second-level ConvLSTM cell, which produces the second-level hidden state $\mathcal{H}_t^2$. This process is repeated for all frames in the context sequence, allowing the final hidden and cell states $(\mathcal{H}_{T_c}^1, \mathcal{C}_{T_c}^1)$ and $(\mathcal{H}_{T_c}^2, \mathcal{C}_{T_c}^2)$ to encapsulate a compressed spatio-temporal representation of the entire observed history.

\subsubsection{Autoregressive Forecasting Decoder}
The decoder's primary task is to generate the future sequence of RMs, $\hat{\mathcal{Y}}$, over $T_p$ time steps in an autoregressive manner. Its architecture mirrors the encoder's but operates in reverse, starting from a compressed representation and progressively reconstructing the full-resolution output. The entire forecasting process is initialized by inheriting the final hidden and cell states from the encoder, which effectively transfers the learned spatio-temporal context of the input sequence to the decoder. Once initialized, the decoder enters a forecasting loop that repeats for each of the $T_p$ future time steps. In each iteration, the second-level (low-resolution) ConvLSTM cell first updates its internal state using its own hidden state from the previous step as input. The resulting feature map is then upsampled by a factor of two via a transposed convolution. To support the reconstruction of spatial details, a skip connection is used: the upsampled feature map is concatenated with the final hidden state from the encoder's first level ($\mathcal{H}_{T_c}^1$). This step injects high-resolution information directly into the decoding path, mitigating information loss from the encoding phase. The resulting fused feature map is then processed by the first-level (high-resolution) decoder ConvLSTM cell. Its output passes through a final $1 \times 1$ convolutional layer with a sigmoid activation function to produce the single-channel RM prediction for the current time step. This process repeats for $T_p$ steps to generate the forecast sequence. The full implementation details, including all hyperparameters, are provided in our open-sourced code repository.

\subsection{Training Objective}
Optimizing a predictive model for RM forecasting requires a loss function that accurately measures the pixel-level discrepancy between the prediction and the ground truth. We employ the Mean Squared Error (MSE), a standard and effective metric for regression-based tasks, as the sole training objective. The MSE loss is calculated as the average of the squared differences between the predicted values and the actual values over all pixels in the sequence as follows.
\begin{equation}
    \mathcal{L}_{MSE} = \frac{1}{T_p \cdot N^2} \sum_{k=1}^{T_p} \sum_{i,j=1}^{N} (\hat{\mathbf{P}}_{t+k}(i,j) - \mathbf{P}_{t+k}(i,j))^2,
\end{equation}
where $\hat{\mathbf{P}}_{t+k}$ is the predicted map at future step $k$, $\mathbf{P}_{t+k}$ is the ground-truth map, and $N \times N$ is the map dimension.

\section{Experiments}
\label{sec:experiments}
We conduct a series of experiments to validate our proposed framework and establish a baseline for the RM prediction task.

\subsection{Experimental Setup}
We use the RadioMapMotion dataset described in Section \ref{sec:dataset}. To conduct a multi-faceted evaluation of our model's predictive capabilities, we employ a comprehensive data splitting strategy designed to disentangle and assess two distinct forms of generalization:
\begin{itemize}
    \item \textbf{Dynamic Generalization:} The model's ability to predict the evolution of the radio environment under novel temporal dynamics, such as previously unseen vehicle trajectories, while operating within a known spatial context like a familiar city map.
    
    \item \textbf{Spatial Generalization:} The model's ability to operate effectively in a completely new spatial context, such as a previously unseen city map, thereby testing its capacity to learn fundamental propagation principles rather than memorize specific layouts.
\end{itemize}

\begin{figure}[t]
    \centering
    \captionsetup{font=small}
    \includegraphics[width=0.9\columnwidth]{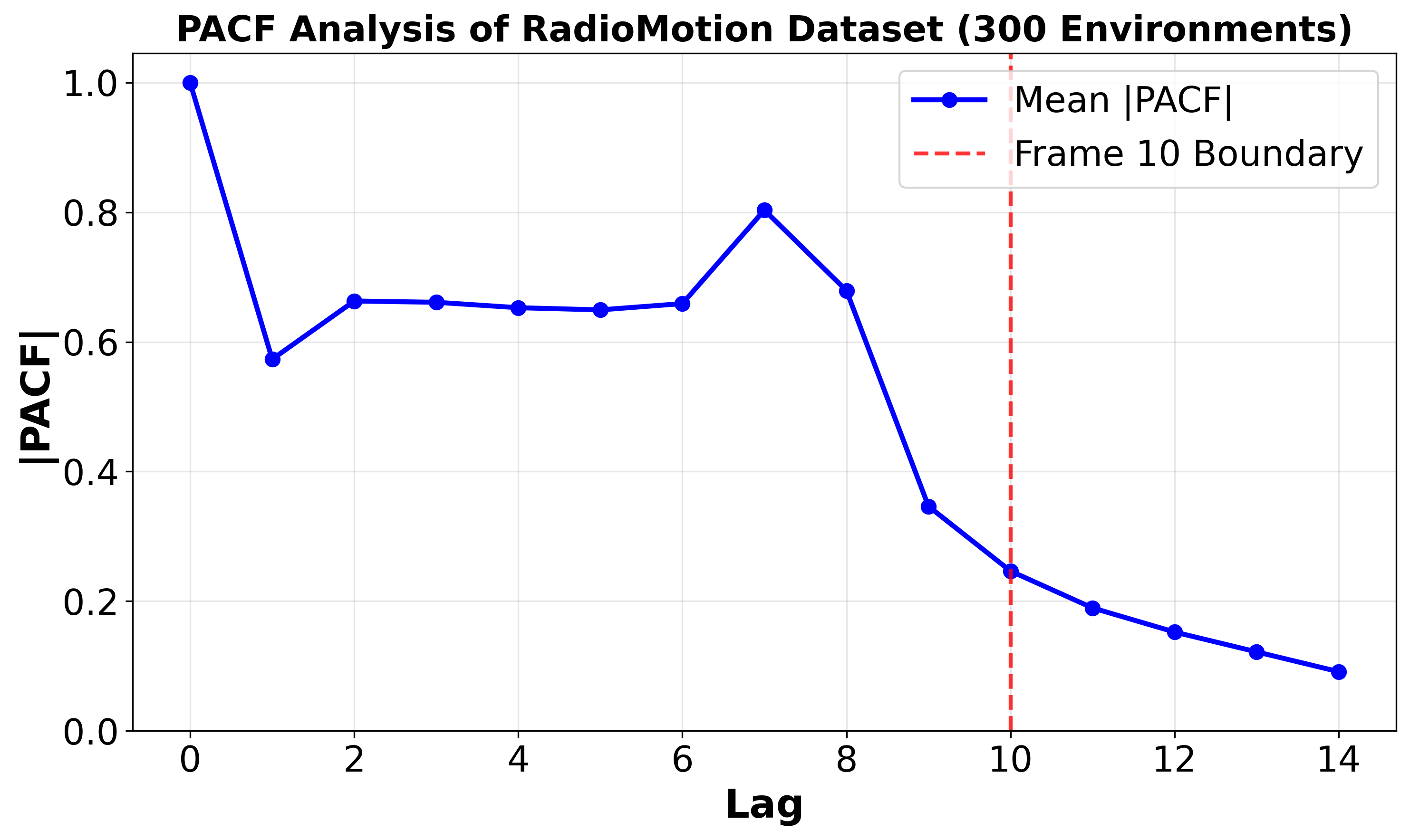} 
    \caption{PACF analysis of the RadioMapMotion dataset, averaged over all sequences.}
    \label{fig:pacf_analysis}
\end{figure}

To facilitate this dual evaluation, our dataset of 300 environments is carefully partitioned. For spatial generalization, we hold out 50 environments entirely, specifically environments 250 through 299, and use all data from them, comprising 5,000 unique sequences, to form Test Set 2, which we refer to as the Unseen Environments test set. This ensures the model is evaluated on geographical layouts it has never encountered during training or validation. The remaining 250 environments, namely environments 0 through 249, are used to construct the training, validation, and dynamic generalization test sets by splitting along the trajectory axis. Sequences from the first three trajectories in each environment form the Training Set, resulting in a total of 15,000 sequences. The fourth trajectory from each environment constitutes the Validation Set with 5,000 sequences, which is used for hyperparameter tuning and model selection. Finally, the fifth trajectory from these same 250 environments, which was not seen during training or validation, forms Test Set 1, known as the Seen Environments but Unseen Trajectories test set, also containing 5,000 sequences. This test set directly measures the model's ability to predict the effects of novel traffic patterns within a familiar geographical context.

For all experiments, we use a context length of $T_c = 10$ frames to predict a future sequence of $T_p = 5$ frames. This choice is empirically validated by our ablation study (Section~\ref{subsec:ablation_context_length}), which demonstrates that $T_c=10$ achieves optimal prediction performance. This finding is supported by Partial Autocorrelation Function (PACF) analysis (Fig.~\ref{fig:pacf_analysis}), which indicates that direct temporal correlations diminish beyond lag 10 ($|\mathrm{PACF}| \approx 0.25$), suggesting it as a reasonable upper bound for capturing relevant dynamics. The prediction horizon $T_p = 5$ (0.5s) provides a time margin for proactive network adaptation in high-mobility scenarios. It should be emphasized that all experiments in this paper are conducted using ground-truth historical RMs from the RadioMapMotion dataset. This design choice allows us to evaluate the pure predictive capability of the models without confounding factors from RM reconstruction errors. In real-world deployments, the quality of the input context sequence $\mathcal{X}$ will depend on the fidelity of the upstream RM reconstruction pipeline, which may affect overall prediction accuracy.

\begin{figure*}[ht]
    \centering
    \captionsetup{font=small, skip=5pt}
    \newlength{\compimgwidth}
    \setlength{\compimgwidth}{0.17\linewidth} 
    \renewcommand{\arraystretch}{1.5} 
    \setlength{\tabcolsep}{3pt}
    \begin{tabular}{@{} c c c c c c @{}}
        \toprule
        \textbf{Model} & \textbf{Frame 11} & \textbf{Frame 12} & \textbf{Frame 13} & \textbf{Frame 14} & \textbf{Frame 15} \\
        \midrule
        
        \rotatebox{90}{\textbf{Last-Frame Repeat}} &
        \adjustbox{width=\compimgwidth}{\includegraphics{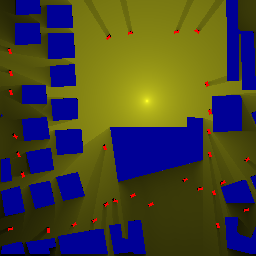}} &
        \adjustbox{width=\compimgwidth}{\includegraphics{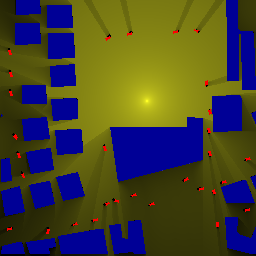}} &
        \adjustbox{width=\compimgwidth}{\includegraphics{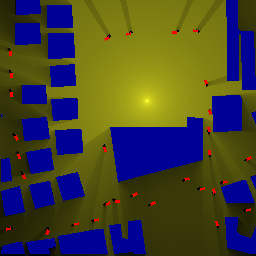}} &
        \adjustbox{width=\compimgwidth}{\includegraphics{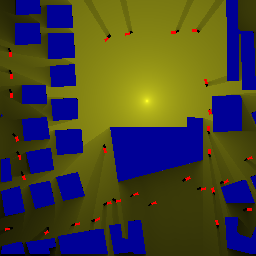}} &
        \adjustbox{width=\compimgwidth}{\includegraphics{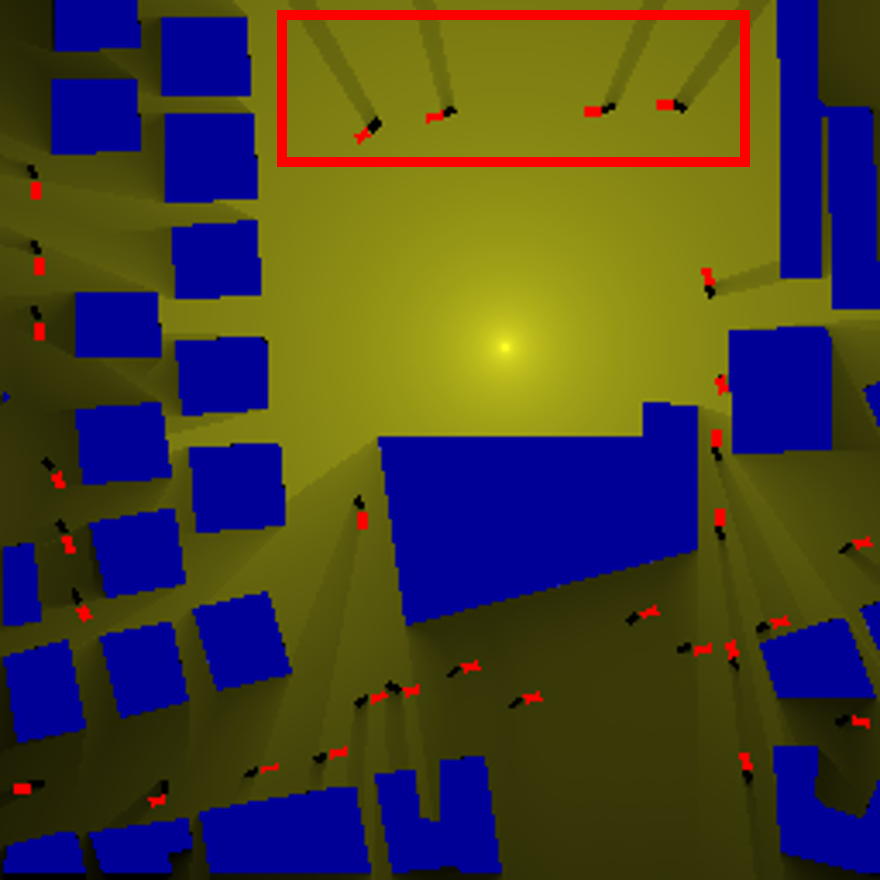}} \\

        \rotatebox{90}{\textbf{Static-UNet}} &
        \adjustbox{width=\compimgwidth}{\includegraphics{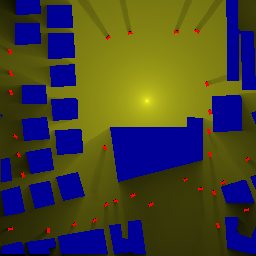}} &
        \adjustbox{width=\compimgwidth}{\includegraphics{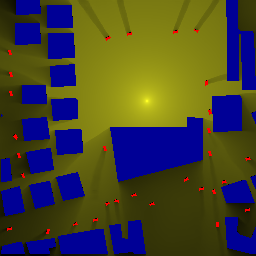}} &
        \adjustbox{width=\compimgwidth}{\includegraphics{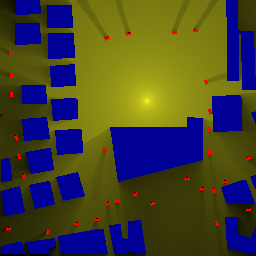}} &
        \adjustbox{width=\compimgwidth}{\includegraphics{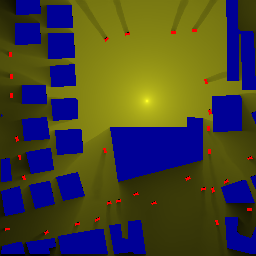}} &
        \adjustbox{width=\compimgwidth}{\includegraphics{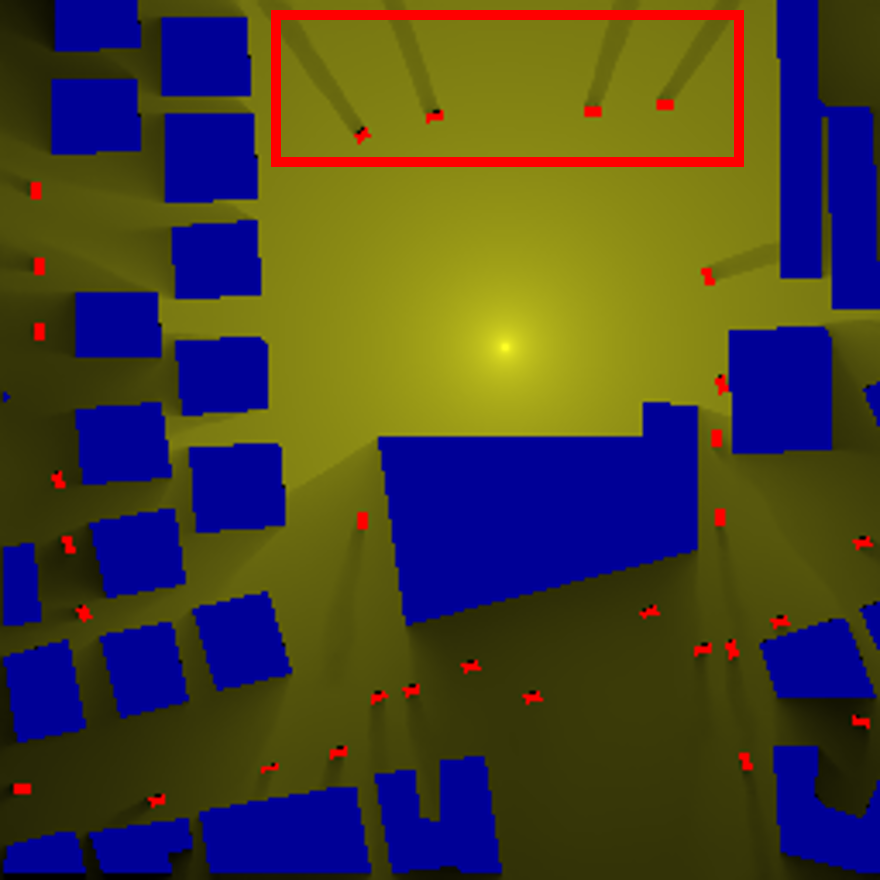}} \\

        \rotatebox{90}{\textbf{MambaUnet}} &
        \adjustbox{width=\compimgwidth}{\includegraphics{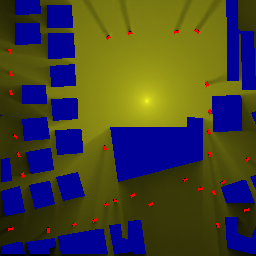}} &
        \adjustbox{width=\compimgwidth}{\includegraphics{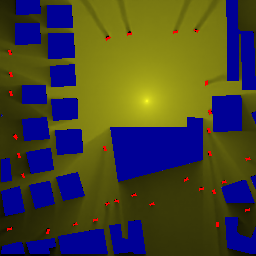}} &
        \adjustbox{width=\compimgwidth}{\includegraphics{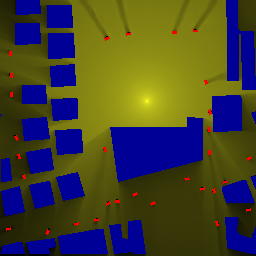}} &
        \adjustbox{width=\compimgwidth}{\includegraphics{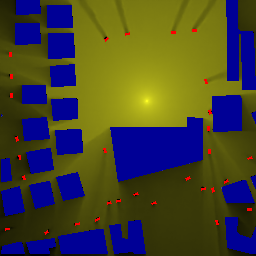}} &
        \adjustbox{width=\compimgwidth}{\includegraphics{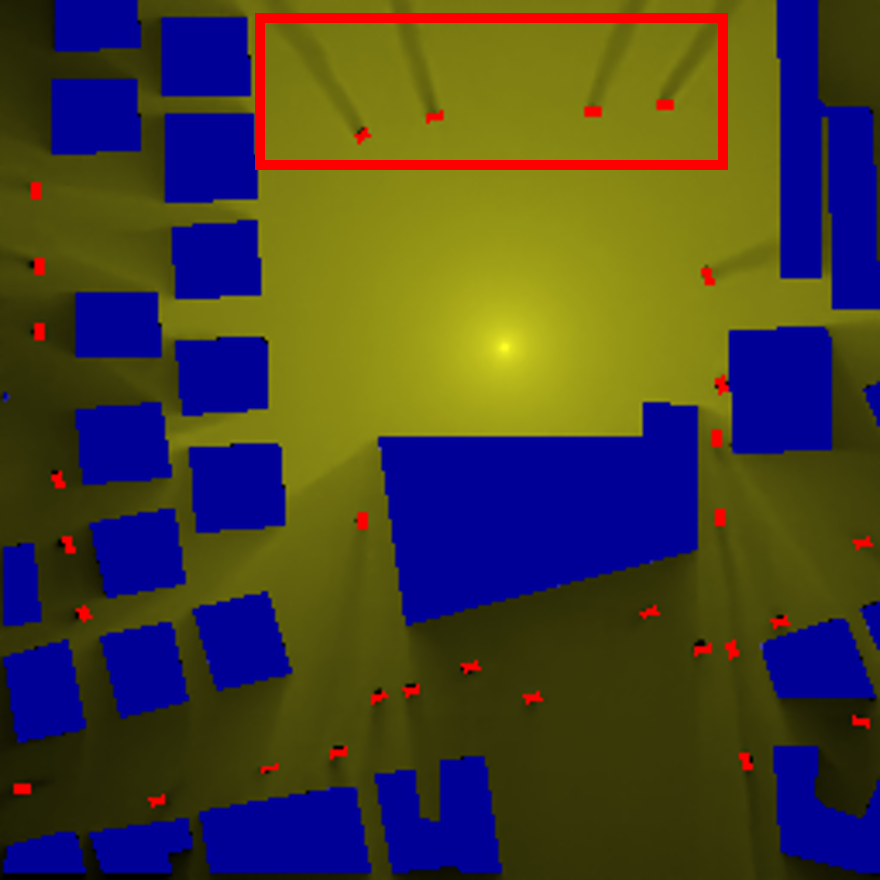}} \\
        
        \rotatebox{90}{\textbf{RadioLSTM(Ours)}} &
        \adjustbox{width=\compimgwidth}{\includegraphics{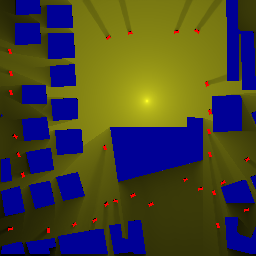}} &
        \adjustbox{width=\compimgwidth}{\includegraphics{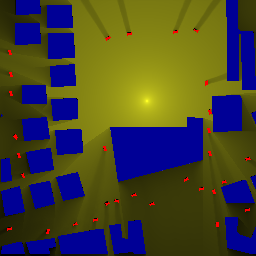}} &
        \adjustbox{width=\compimgwidth}{\includegraphics{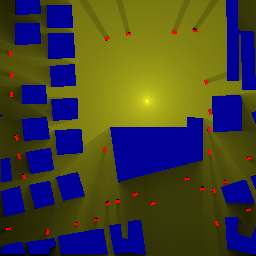}} &
        \adjustbox{width=\compimgwidth}{\includegraphics{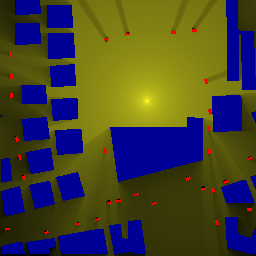}} &
        \adjustbox{width=\compimgwidth}{\includegraphics{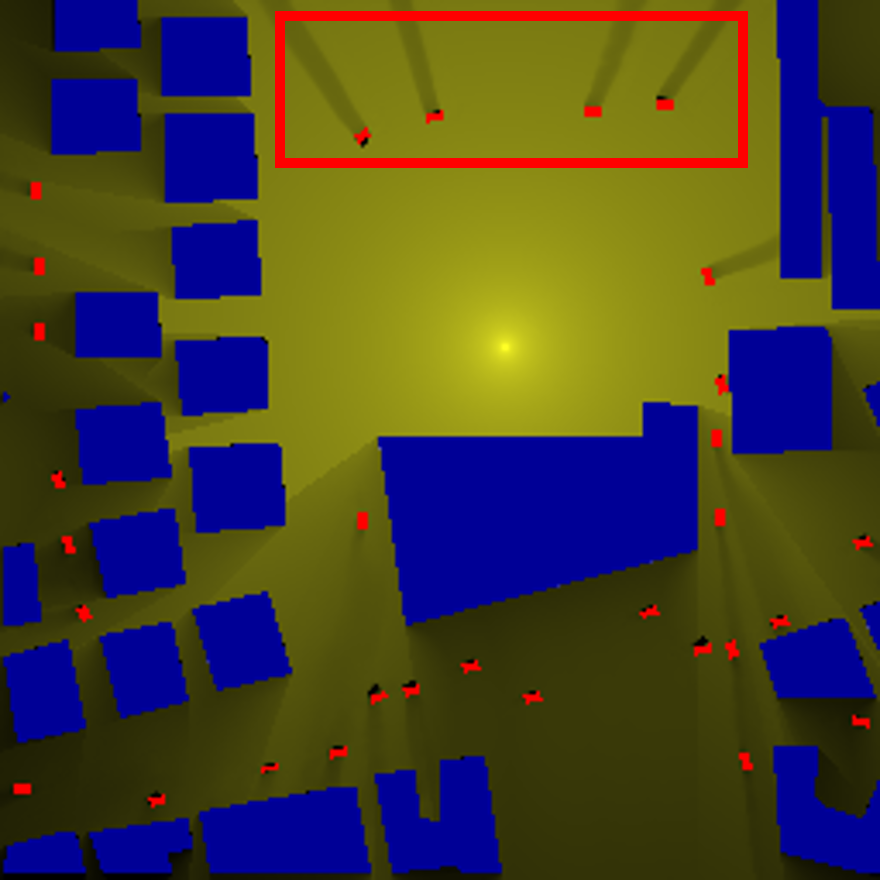}} \\

        \rotatebox{90}{\textbf{Ground Truth}} &
        \adjustbox{width=\compimgwidth}{\includegraphics{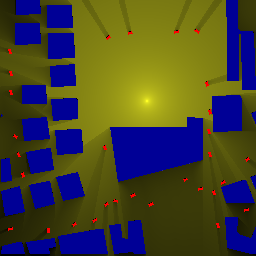}} &
        \adjustbox{width=\compimgwidth}{\includegraphics{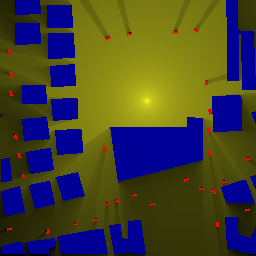}} &
        \adjustbox{width=\compimgwidth}{\includegraphics{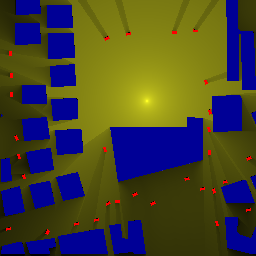}} &
        \adjustbox{width=\compimgwidth}{\includegraphics{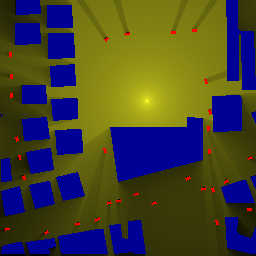}} &
        \adjustbox{width=\compimgwidth}{\includegraphics{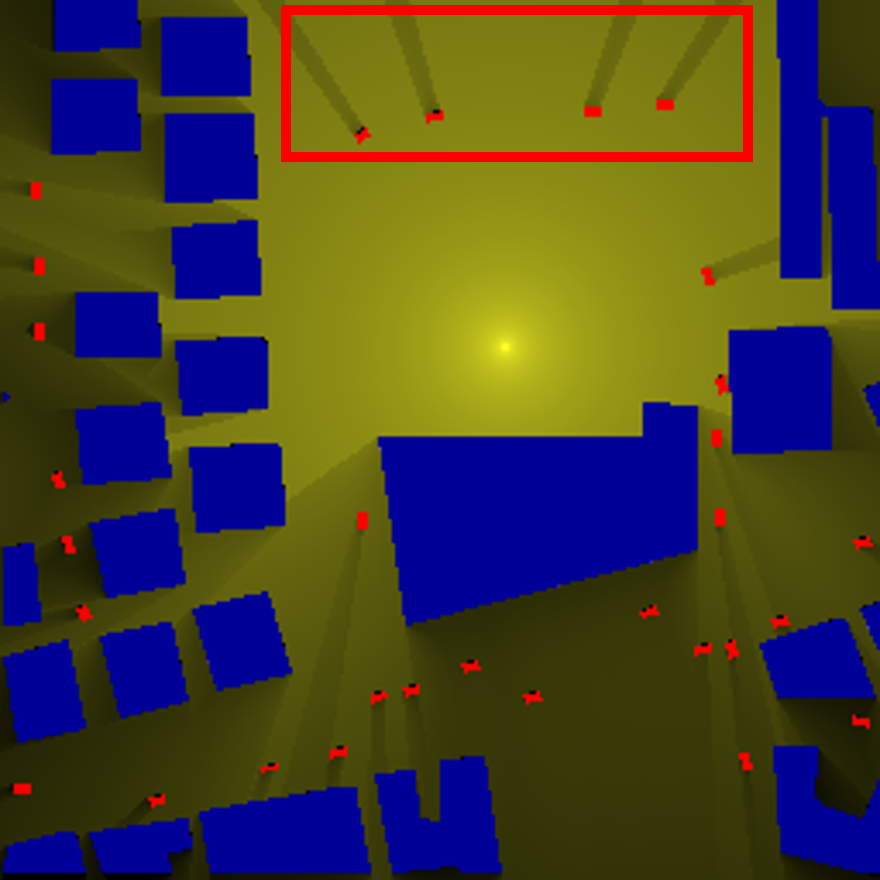}} \\
        \bottomrule
    \end{tabular}
    
    \caption{Qualitative comparison for dynamic generalization (seen environment). Each row illustrates the output of a specific method for an unseen vehicle trajectory within a familiar environment.}
    \label{fig:qualitative_comparison_seen}
\end{figure*}

\begin{figure*}[ht]
    \centering
    \captionsetup{font=small, skip=5pt}
    \setlength{\compimgwidth}{0.17\linewidth} 
    \renewcommand{\arraystretch}{1.0}
    \setlength{\tabcolsep}{3pt}
    \begin{tabular}{@{} c c c c c c @{}}
        \toprule
        \textbf{Model} & \textbf{Frame 11} & \textbf{Frame 12} & \textbf{Frame 13} & \textbf{Frame 14} & \textbf{Frame 15} \\
        \midrule
        
        \rotatebox{90}{\textbf{Last-Frame Repeat}} &
        \adjustbox{width=\compimgwidth}{\includegraphics{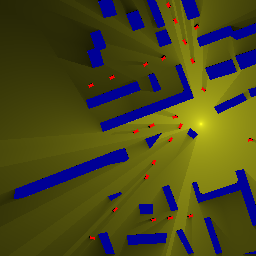}} &
        \adjustbox{width=\compimgwidth}{\includegraphics{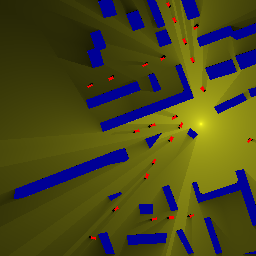}} &
        \adjustbox{width=\compimgwidth}{\includegraphics{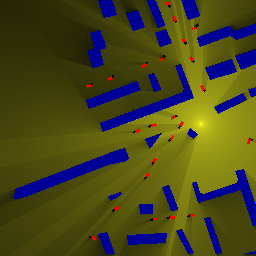}} &
        \adjustbox{width=\compimgwidth}{\includegraphics{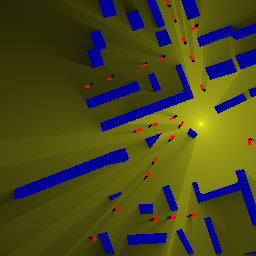}} &
        \adjustbox{width=\compimgwidth}{\includegraphics{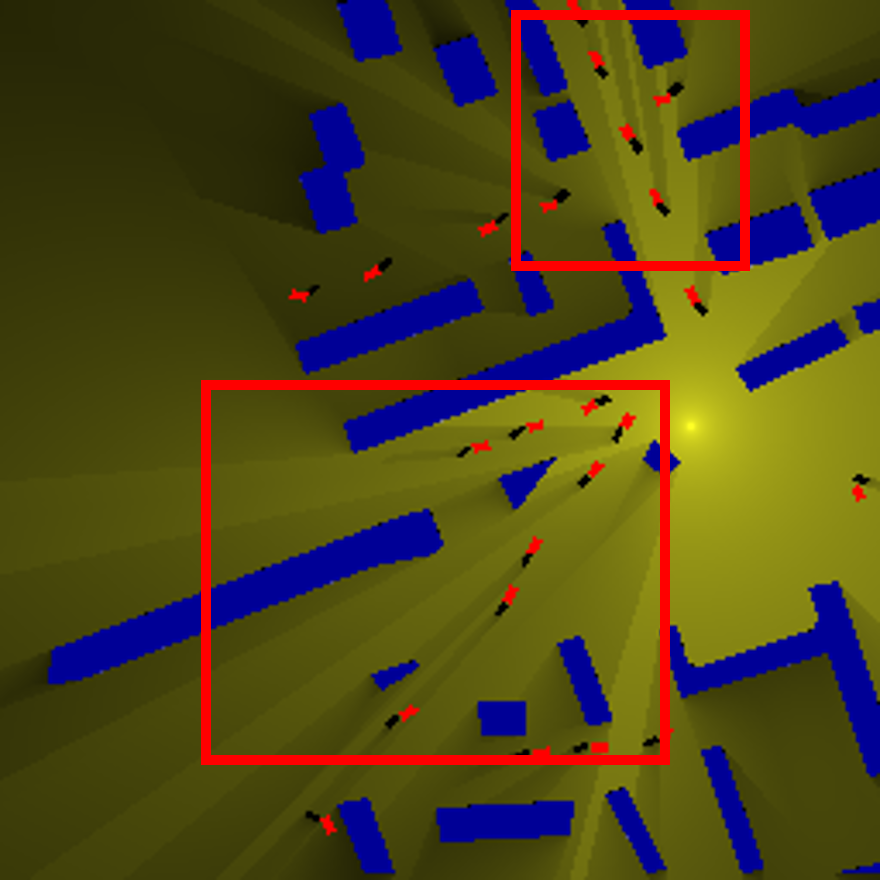}} \\

        \rotatebox{90}{\textbf{Static-UNet}} &
        \adjustbox{width=\compimgwidth}{\includegraphics{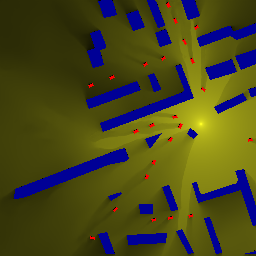}} &
        \adjustbox{width=\compimgwidth}{\includegraphics{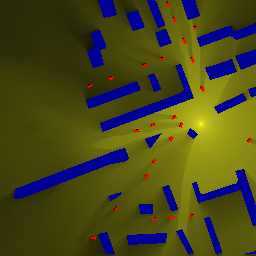}} &
        \adjustbox{width=\compimgwidth}{\includegraphics{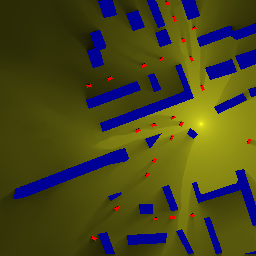}} &
        \adjustbox{width=\compimgwidth}{\includegraphics{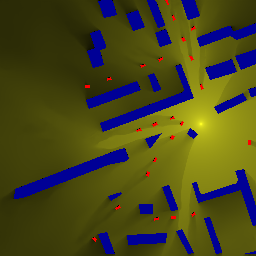}} &
        \adjustbox{width=\compimgwidth}{\includegraphics{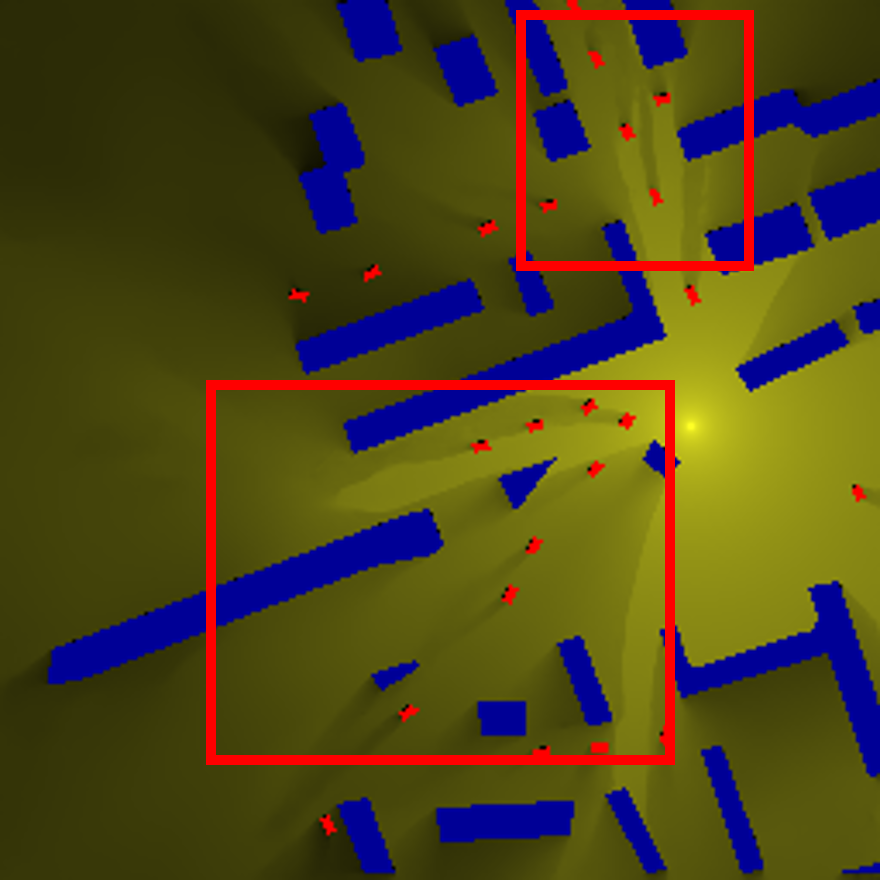}} \\

        \rotatebox{90}{\textbf{MambaUnet}} &
        \adjustbox{width=\compimgwidth}{\includegraphics{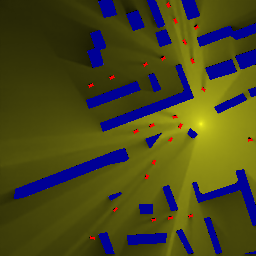}} &
        \adjustbox{width=\compimgwidth}{\includegraphics{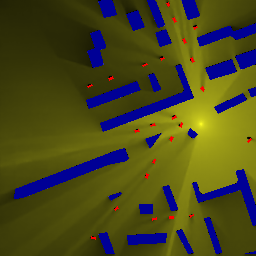}} &
        \adjustbox{width=\compimgwidth}{\includegraphics{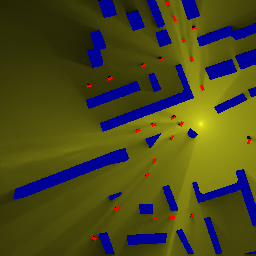}} &
        \adjustbox{width=\compimgwidth}{\includegraphics{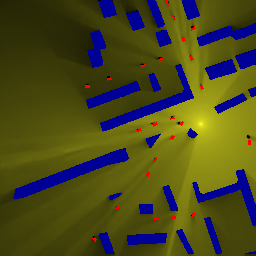}} &
        \adjustbox{width=\compimgwidth}{\includegraphics{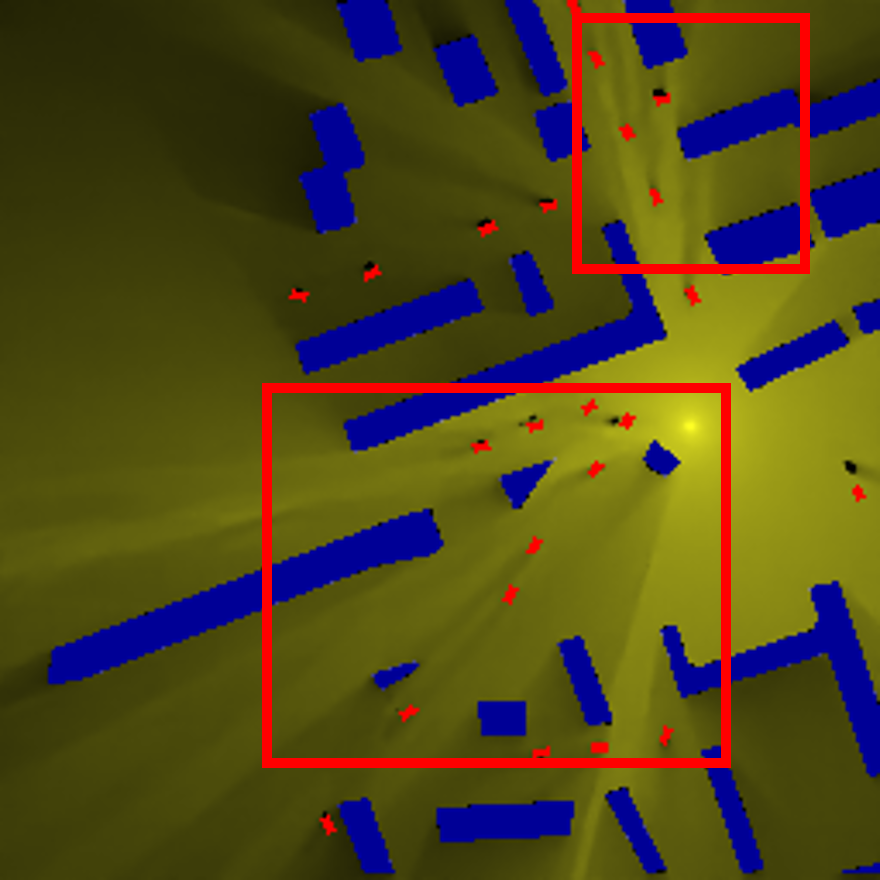}} \\
        
        \rotatebox{90}{\textbf{RadioLSTM(Ours)}} &
        \adjustbox{width=\compimgwidth}{\includegraphics{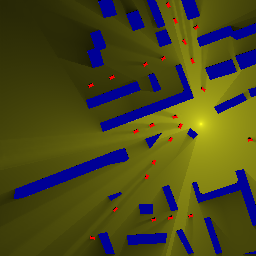}} &
        \adjustbox{width=\compimgwidth}{\includegraphics{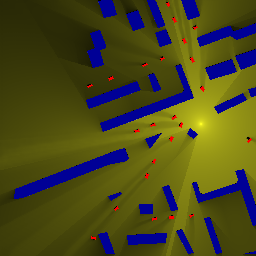}} &
        \adjustbox{width=\compimgwidth}{\includegraphics{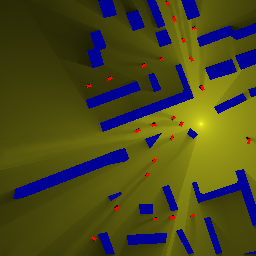}} &
        \adjustbox{width=\compimgwidth}{\includegraphics{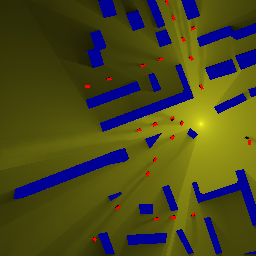}} &
        \adjustbox{width=\compimgwidth}{\includegraphics{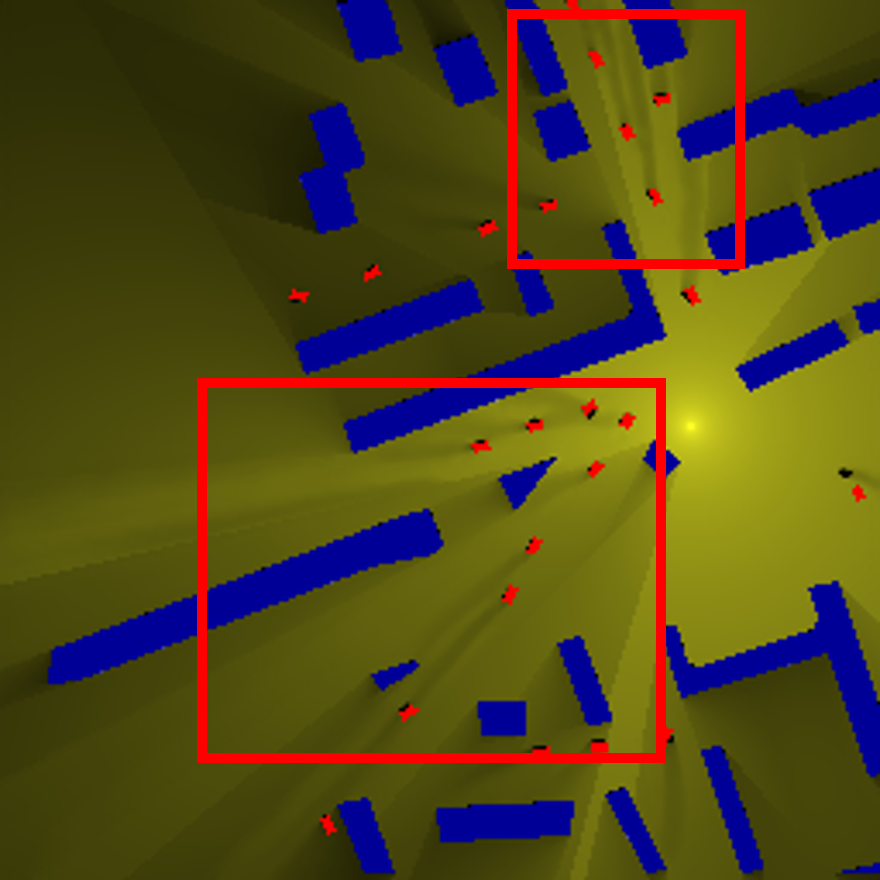}} \\

        \rotatebox{90}{\textbf{Ground Truth}} &
        \adjustbox{width=\compimgwidth}{\includegraphics{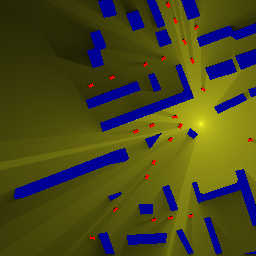}} &
        \adjustbox{width=\compimgwidth}{\includegraphics{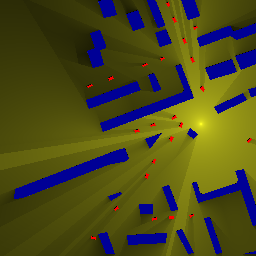}} &
        \adjustbox{width=\compimgwidth}{\includegraphics{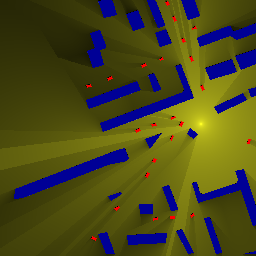}} &
        \adjustbox{width=\compimgwidth}{\includegraphics{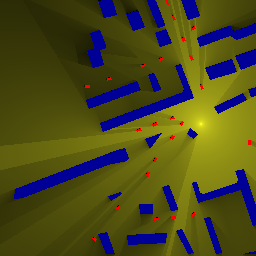}} &
        \adjustbox{width=\compimgwidth}{\includegraphics{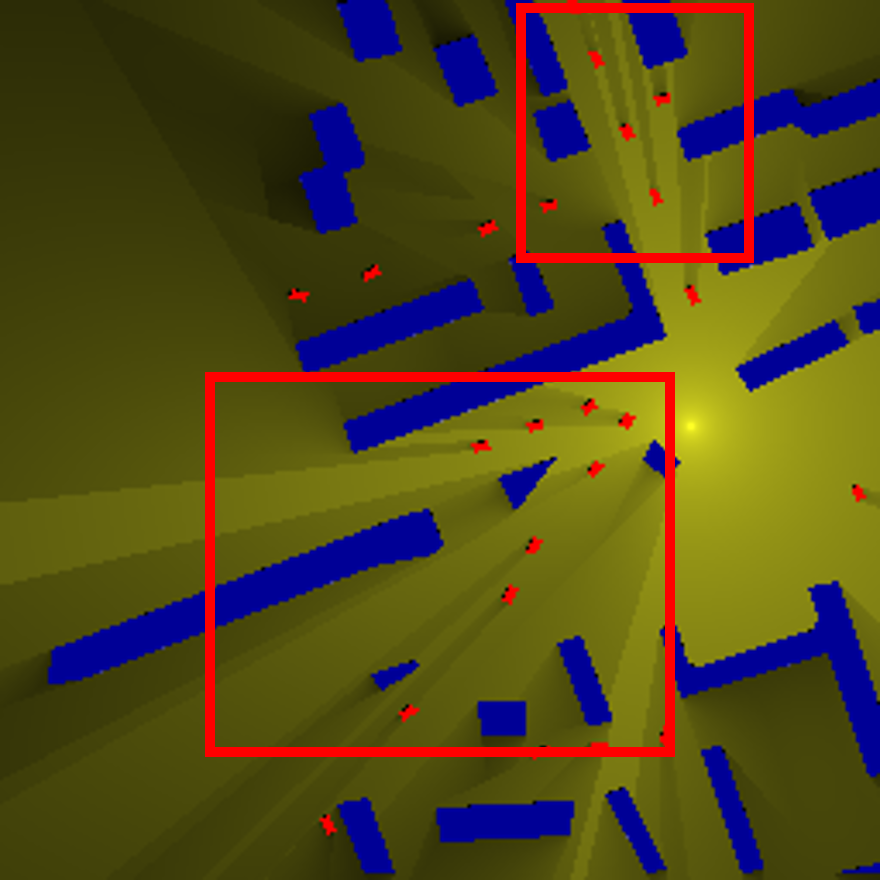}} \\
        \bottomrule
    \end{tabular}
    \caption{Qualitative comparison for spatial generalization (unseen environment). This figure shows the models' performance in a completely new environment not encountered during training.}
    \label{fig:qualitative_comparison_unseen}
\end{figure*}

\subsection{Baselines}
To validate our proposed framework, we compare RadioLSTM against several carefully chosen baselines, each designed to test a specific aspect of our core hypothesis.

\begin{itemize}
    \item \textbf{Last-Frame Repeat}: A naive, non-learning baseline that simply repeats the last observed frame, $\mathbf{P}_t$, for all future time steps. It serves as a fundamental lower bound, quantifying the environment's dynamism and the necessity of temporal modeling.

    \item \textbf{Static-UNet (Frame-by-Frame)}: This baseline adapts a powerful static model like RadioUNet~\cite{levie2021radiounet} to a sequence task by training it to predict only the next frame ($\mathbf{P}_{t+1}$) from the current one ($\mathbf{P}_t$). The full sequence is then generated autoregressively. This comparison directly tests our central claim that explicitly modeling long-term temporal dependencies is superior to treating the problem as a series of independent, frame-to-frame transitions.

    \item \textbf{ConvMambaUNet}: A modern baseline that integrates convolutional spatial feature extraction with Mamba-based temporal modeling in a UNet framework. Specifically, it employs a hybrid \textit{ConvMambaBlock} that first applies spatial convolutions, then sequences spatial patches into a temporal stream processed by Mamba\cite{zhang2025convmamba}. The architecture follows an encoder-decoder UNet structure with skip connections and autoregressive forecasting, designed to balance spatial fidelity and temporal coherence. This model tests whether SSMs can be effectively adapted to high-resolution, detail-sensitive radio map prediction.
\end{itemize}

More advanced video prediction architectures, such as those based on Transformers or diffusion models, have demonstrated state-of-the-art performance in general video synthesis, we prioritized more efficient baselines. This choice is motivated by the low-latency inference requirement of proactive network management and the physical nature of our task. The evolution of radio maps is governed by strong spatio-temporal locality, a property that convolutional recurrent networks are inherently suited to capture efficiently, while more complex models often introduce significant computational overhead. As our experimental results will later show, even an architecture like ConvMambaUNet, which is celebrated for its efficiency, is outperformed by our simpler recurrent model in terms of inference speed. Thus, the selected baselines are sufficient to serve the foundational purpose of this work, leaving a broader exploration of the accuracy-efficiency trade-off with more complex generative models for future research.

\subsection{Implementation and Metrics}
The model is implemented in PyTorch and trained using PyTorch Lightning on a single NVIDIA A40 GPU. We use the AdamW optimizer with a learning rate of $1\times10^{-3}$ and mixed-precision training to enhance efficiency. Training runs for up to 1000 epochs with early stopping if validation loss plateaus for 30 epochs, and the best checkpoint is selected based on minimum validation loss. The batch size is 16.

Performance is evaluated using standard metrics computed on the normalized pathloss values. Prediction accuracy is measured by Normalized Mean Square Error (NMSE) and Root Mean Square Error (RMSE), defined as follows.
\begin{equation}
    \text{NMSE} = \frac{\sum_{m=1}^{M}\sum_{n=1}^{N}(I_{\text{gt}}(m,n) - I_{\text{pred}}(m,n))^2}{\sum_{m=1}^{M}\sum_{n=1}^{N}I_{\text{gt}}^2(m,n)},
\end{equation}
\begin{equation}
    \text{RMSE} = \sqrt{\frac{1}{M N} \sum_{m=1}^{M}\sum_{n=1}^{N}(I_{\text{gt}}(m,n) - I_{\text{pred}}(m,n))^2}.
\end{equation}
Structural fidelity is assessed using the Structural Similarity Index Measure (SSIM) and Peak Signal-to-Noise Ratio (PSNR), defined as follows.
\begin{equation}
    \text{SSIM}(x,y) = l(x,y)^\alpha \cdot c(x,y)^\beta \cdot s(x,y)^\gamma,
\end{equation}
\begin{equation}
    \text{PSNR} = 10 \cdot \log_{10}\left(\frac{{\text{MAX}_I}^2}{\text{MSE}}\right),
\end{equation}
where $\text{MAX}_I$ is 1.0 for our normalized data. Lower NMSE and RMSE, and higher SSIM and PSNR, indicate better performance.

\begin{table*}[ht]
    \centering
    \captionsetup{font=small}
    \caption{Quantitative Comparison for RM Prediction ($T_c=10, T_p=5$). The evaluation is performed on two test sets: Test Set 1 for dynamic generalization (seen environments, unseen trajectories) and Test Set 2 for spatial generalization (unseen environments). The best performance in each category is highlighted in \ourmodelval{red}, and the second best is \secondbestval{underlined in blue}.}
    \label{tab:quantitative_results}
    \renewcommand{\arraystretch}{1.1}
    \begin{adjustbox}{width=0.8\textwidth, center}
        \begin{tabular}{l l c c c c}
            \toprule
            \textbf{Test Scenario} & \textbf{Model} & \textbf{NMSE} $\downarrow$ & \textbf{RMSE} $\downarrow$ & \textbf{SSIM} $\uparrow$ & \textbf{PSNR (dB)} $\uparrow$ \\
            \midrule
            \multirow{4}{*}{\begin{tabular}[c]{@{}l@{}}\textbf{Test Set 1} \\ (Dynamic Generalization)\end{tabular}}
            & Last-Frame Repeat & 0.009517 & 0.029257 & 0.9421 & 30.6755 \\
            & Static-UNet & \secondbestval{0.002751} & \secondbestval{0.015731} & \secondbestval{0.9700} & \secondbestval{36.0648} \\
            & ConvMambaUNet & 0.004300 & 0.019666 & 0.9656 & 34.5753 \\
            & \textbf{RadioLSTM(Ours)} & \ourmodelval{0.002562} & \ourmodelval{0.015181} & \ourmodelval{0.9779} & \ourmodelval{36.3742} \\
            \cmidrule(lr){1-6}
            \multirow{4}{*}{\begin{tabular}[c]{@{}l@{}}\textbf{Test Set 2} \\ (Spatial Generalization)\end{tabular}}
            & Last-Frame Repeat & 0.007519 & 0.028530 & 0.9460 & 30.8941 \\
            & Static-UNet & 0.011265 & 0.034921 & 0.9338 & 29.1382 \\
            & ConvMambaUNet & \secondbestval{0.004553} & \secondbestval{0.022201} & \secondbestval{0.9609} & \secondbestval{33.0726} \\
            & \textbf{RadioLSTM(Ours)} & \ourmodelval{0.002121} & \ourmodelval{0.015154} & \ourmodelval{0.9800} & \ourmodelval{36.3894} \\
            \bottomrule
        \end{tabular}
    \end{adjustbox}
\end{table*}

\begin{table}[ht]
    \centering
    \captionsetup{font=small}
    \caption{Efficiency Comparison.}
    \label{tab:inference_efficiency}
    \renewcommand{\arraystretch}{1.1}
    \begin{tabular*}{0.8\linewidth}{l c c}
        \toprule
        \textbf{Model} & \makecell[c]{\textbf{Inference time} \\ \textbf{(ms/frame)} $\downarrow$} & \makecell[c]{\textbf{GPU memory} \\ \textbf{(MB)} $\downarrow$} \\
        \midrule
        Static-UNet  & 3.9 & \ourmodelval{525} \\
        ConvMambaUNet & 4.9 & 2468 \\
        \textbf{RadioLSTM(Ours)} & \ourmodelval{1.3} & \secondbestval{927} \\
        \bottomrule
    \end{tabular*}
\end{table}
\subsection{Performance Evaluation and Analysis}
\subsubsection{Quantitative Results}
Table~\ref{tab:quantitative_results} presents the quantitative results, which highlight the superiority of spatio-temporal modeling across two distinct generalization scenarios.

On Test Set 1, which evaluates dynamic generalization, RadioLSTM achieves the best performance across all metrics. This confirms that explicitly modeling temporal dependencies yields more accurate predictions than frame-by-frame approaches such as Static-UNet, even within familiar environments. The advantages of our approach are most pronounced on Test Set 2, which evaluates spatial generalization. In this setting, the performance of Static-UNet degrades sharply; for example, its NMSE increases by over 300 percent, indicating that it overfits to training layouts and fails to learn generalizable propagation principles. In contrast, both temporal models generalize effectively to unseen geographies. Although ConvMambaUNet demonstrates robust performance, RadioLSTM again secures the top position with stable and high-fidelity predictions. This suggests that by learning the fundamental dynamics of radio wave propagation, RadioLSTM can robustly generalize to entirely new environments.

Furthermore, efficiency is critical for proactive network control. As detailed in Table~\ref{tab:inference_efficiency}, RadioLSTM excels in inference speed, achieving an average of 1.3 milliseconds per frame, which is significantly faster than Static-UNet at 3.9 milliseconds and ConvMambaUNet at 4.9 milliseconds. Although Static-UNet is the most memory-efficient, its accuracy is limited, especially in unseen environments. Conversely, while ConvMambaUNet offers competitive accuracy, its high computational and memory demands may hinder real-time deployment. Therefore, RadioLSTM presents the most compelling trade-off, delivering superior accuracy and the lowest latency with a moderate memory footprint, establishing it as a prime candidate for real-time proactive applications.

\subsubsection{Qualitative Results}
This section provides a visual assessment of model performance, with qualitative comparisons for dynamic and spatial generalization presented in Fig.~\ref{fig:qualitative_comparison_seen} and Fig.~\ref{fig:qualitative_comparison_unseen}, respectively.

In the dynamic generalization test within a familiar environment (Fig.~\ref{fig:qualitative_comparison_seen}), the Last-Frame Repeat baseline produces a static output that does not reflect vehicle motion. This results in a visible misalignment between the predicted pathloss shadow and the actual vehicle position in subsequent frames. Both RadioLSTM and Static-UNet generate outputs that show movement consistent with the ground truth, indicating they capture the temporal evolution to some degree. The ConvMambaUNet also generates moving predictions, though its outputs exhibit slightly smoother transitions and less distinct boundaries around vehicle-induced shadows compared to RadioLSTM. In the spatial generalization test on an unseen environment (Fig.~\ref{fig:qualitative_comparison_unseen}), the Static-UNet's predictions show reduced structural clarity, particularly in complex building layouts. The ConvMambaUNet maintains overall structure but displays minor artifacts in regions of high propagation complexity. In contrast, RadioLSTM’s predictions preserve sharper edges and finer details, showing closer alignment with the ground truth across all five predicted frames. This visual observation is consistent with the quantitative metrics reported in Table~\ref{tab:quantitative_results}.

\begin{figure*}[ht]
    \centering
    \captionsetup{font=small}
    \includegraphics[width=0.95\textwidth]{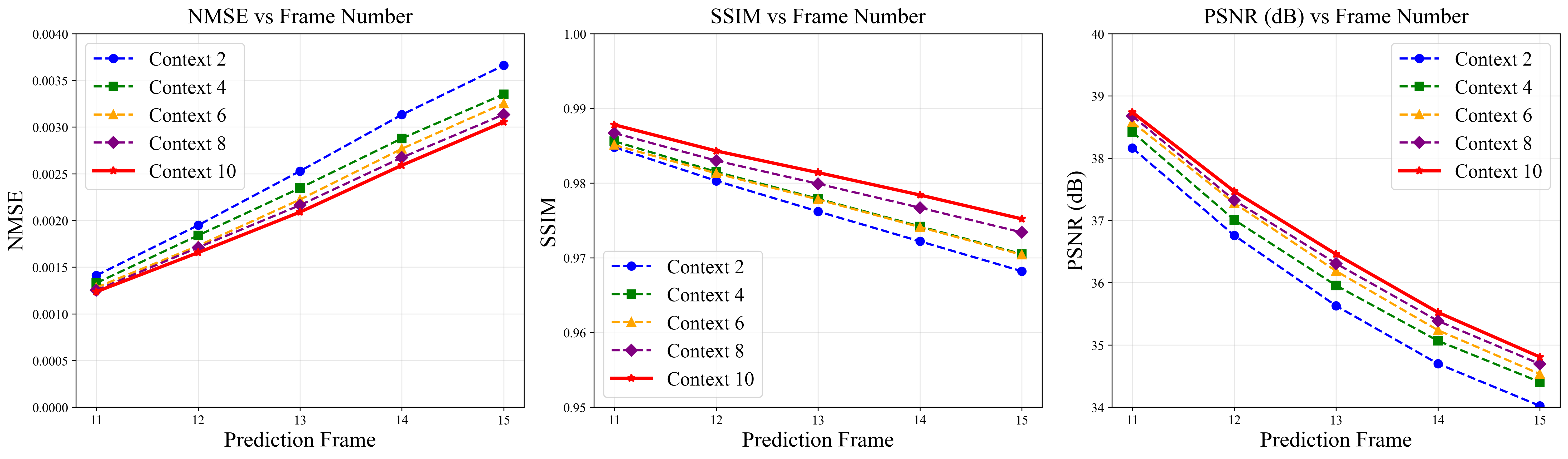}
    \caption{Frame-wise performance comparison of RadioLSTM with different context lengths ($T_c$) on Test Set 2. The context length $T_c$ refers to the number of historical radio map frames used by the model to predict the last 5 frames. For example, if $T_c=4$, the model uses Frames 6-9 as input to predict Frames 11-15.}
    \label{fig:ablation_context}
\end{figure*}

\subsubsection{Analysis of Temporal Modeling Choice}
The task of spatio-temporal radio map prediction primarily involves modeling fine-grained changes caused by moving vehicles, which manifest as localized variations in pathloss shadows across consecutive frames. Convolutional operations are well suited to capture such local spatial patterns, and when combined with recurrent gating in ConvLSTM, they effectively model short-term temporal dynamics. In contrast, the Mamba architecture is designed to handle long-range dependencies in sequences with linear computational complexity. However, the input sequences in this work consist of only 15 frames, with a context length of 10 and a prediction horizon of 5, which constitutes a short-sequence regime where the efficiency advantage of Mamba is not fully realized. Moreover, the radio environment is dominated by static structures such as buildings, and the dynamic component occupies a relatively small spatial region. Under these conditions, global sequence modeling offers limited benefit, whereas local spatio-temporal modeling preserves high-frequency details critical for pathloss fidelity. Although more complex architectures such as Transformers or diffusion models may achieve marginally higher accuracy, they incur significantly higher computational cost. Given that the current model already achieves high structural similarity and operates at low latency, the trade-off favors the simpler ConvLSTM-based design for real-time deployment.

\subsection{Ablation Study}
\label{subsec:ablation_context_length}
To validate the choice of context length $T_c = 10$ and investigate its impact on prediction accuracy, we conducted an ablation study by training RadioLSTM with varying context lengths ($T_c = 2, 4, 6, 8, 10$) while keeping the prediction horizon fixed at $T_p = 5$. The evaluation was performed on Test Set 2 (unseen environments) to assess generalization capability. The results are visualized in Fig.~\ref{fig:ablation_context}. Two consistent patterns emerge from the analysis.

First, regardless of the context length, all models exhibit a systematic performance decay across the prediction horizon. Metrics such as NMSE and RMSE show a monotonic increase from the first to the fifth predicted frame, SSIM and PSNR correspondingly decrease. This behavior is characteristic of autoregressive forecasting systems, where prediction errors accumulate over time due to the recursive nature of the generation process: each subsequent frame is conditioned on the model’s own previous output. Second, increasing the context length consistently improves prediction fidelity, with $T_c=10$ yielding the best overall performance. This improvement is most pronounced in later prediction steps, where longer contexts help mitigate error accumulation by providing a more accurate initial spatio-temporal state for the decoder. It is worth noting that even with a minimal context of $T_c=2$, the model maintains reasonable performance. This is because the encoder, regardless of input sequence length, compresses the observed history into a single latent state that initializes the forecasting process. A sufficiently expressive encoder can extract meaningful dynamics even from short sequences, resulting in a moderate performance gap between $T_c=2$ and $T_c=10$. Nevertheless, the quantitative advantage of a longer context confirms that $T_c=10$ captures a more complete representation of the environment’s temporal dynamics, leading to more robust and accurate multi-step forecasts. Addressing the challenge of even longer prediction horizons (e.g., $T_p > 10$) remains an important open question, as error accumulation is likely to become more severe. Future work could involve extending the RadioMapMotion dataset with longer continuous sequences to specifically investigate and develop models capable of robust long-term forecasting.

\section{Conclusion and Future Work}
\label{sec:conclusion}
This paper has introduced the task of spatio-temporal RM prediction, which aims to forecast future radio conditions based on historical observations. To support research in this area, we present RadioMapMotion, a dataset that captures the continuous evolution of radio environments driven by vehicle motion. We also propose RadioLSTM, a ConvLSTM-based architecture, as a baseline method for this task. Experimental results indicate that RadioLSTM achieves lower prediction error and higher structural similarity compared to evaluated baselines, particularly when tested on totally new environments. The model also demonstrates low inference latency, which is a practical consideration for real-time applications. Future work will focus on extending the dataset with more diverse environmental conditions, carrier frequencies, and realistic traffic dynamics, while also exploring advanced prediction architectures and robust integration with real-world, noisy radio measurements.

\bibliographystyle{IEEEtran}
\bibliography{ref} 

\end{document}